# Creation of high energy/intensity bremsstrahlung by a multitarget and focusing of the scattered electrons by small-angle backscatter at a cone wall and a magnetic field II - Enhancement of the outcome of linear accelerators in radiotherapy

W. Ulmer[1,2]

[1]Klinikum München-Pasing, Dept. of Radiotherapy and [2]MPI of Biophysical Chemistry, Göttingen, Germany

*E-Mail:* waldemar.ulmer@gmx.net

**Abstract**

The yield of bremsstrahlung from collisions of fast electrons (energy at least 6 MeV) with a Tungsten target can be significantly improved by exploitation of Tungsten wall scatter in a multi-layered target. A simplified version of a previously developed principle is also able to focus small angle scattered electrons by a Tungsten wall. It is necessary that the thickness of each Tungsten layer does not exceed 0.04 mm - a thickness of 0.03 mm is suitable for accelerators in medical physics. Further focusing of electrons results from suitable magnetic fields with field strength between 0.5 Tesla and 1.2 Tesla (if the cone with multi-layered targets is rather narrow). Linear accelerators in radiation therapy only need focusing by wall scatter without further magnetic fields (standard case: 31 plates with 0.03 mm thickness and 1 mm distance between the plates). We considered three cases with importance in medical physics: A very small cone with additional magnetic field for focusing (field diameter at 90 cm depth: 6cm), a medium cone with optional magnetic field (field diameter at 90 cm depth: 13 cm) and broad cone without magnetic field (field diameter at 90 cm depth: 30 cm). All these cases can be positioned in a carousel. Measurements have been performed in the existing carousel positioned in the plane of the flattening filter and scatter foils for electrons.

## 1. Introduction

A principal problem in the creation of bremsstrahlung of linear accelerators used in radiotherapy is the lack of efficiency, since only a rather small part of the created bremsstrahlung is available for applications. Even for a 40 x 40 $cm^2$ field (distance of 100 cm from focus) the bremsstrahlung yield is small, and most of it goes lost at the primary collimator and jaws. We have usually to deal in radiotherapeutic applications with much smaller field sizes than 40 x 40 $cm^2$. Thus the effectivity in IMRT and stereotaxy is much smaller. A further source of bremsstrahlung loss is the flattening filter, which has to homogenize the transverse profile. Brahme and Svensson (1996) suggested a linear accelerator using a multiple Beryllium target, and electron energies of the order of 80 – 100 MeV to yield an energy spectrum comparable to a conventional machine with 4 – 6 MeV with a single Tungsten target. However, this concept has the disadvantage that electrons decelerated down to ca. 45 – 50 MeV have to be removed by a magnetic field, since they would produce rather low bremsstrahlung energy in further Beryllium targets.

In this communication, we present three modified configurations of a multitarget system of very thin Tungsten layers by theoretical and experimental results, i.e. the thickness is significantly smaller than 0.1 mm to create bremsstrahlung in a much more efficient way by the ordinarily used electron energies between 6 and 20 MeV.

This new way of bremsstrahlung creation exploits two physical effects, which can be used for focusing of scattered electrons, namely the wall scatter of high Z materials such as Tungsten and, a further option, a suitable external magnetic field, which acts as a magnetic lense.

## 2. Material and methods

### 2.1. Schematic representation of bremsstrahlung by a linear accelerator

The following figure 1 shows the essential component modules of a linear accelerator. It starts with the impinging electron current on the bremsstrahlung target (usually Tungsten). The foregoing modules of the beam line such as klystron/magnetron, modulator, acceleration tube, deflection of the electron current by bending magnet are not of importance here. Thus in the target two competition processes occur, namely bremsstrahlung creation and multiple scatter of electrons by simultaneous production of heat.

This multiple scatter and heat production must be regarded as the reason, that the bremsstrahlung creation does not show any preference direction. The beam line according to Figure 1 indicates that only that part of the bremsstrahlung can be used, which can pass through the opening of the primary collimator. The flattening filter immediately below the vacuum window affects the shape of the beam, which is further controlled by the jaws to obtain the desired field size. However, the flattening filter significantly attenuates the intensity of the photon beam, and due to the inevitable Compton scattering, it also acts as a second source, which affects the shape of the profiles of larger field sizes (e.g. the penumbra).

As a resume we can conclude that the present linear accelerators do not provide a high efficiency, in particular with regard to the novel irradiation techniques such as Stereotaxy, Rapid Arc, and IMRT, where very small field sizes are required and most of the produced bremsstrahlung goes lost by shielding of the accelerator head.

Therefore the question arises, in which way we can significantly improve the yield of bremsstrahlung and reduce the required shielding material in the accelerator head.

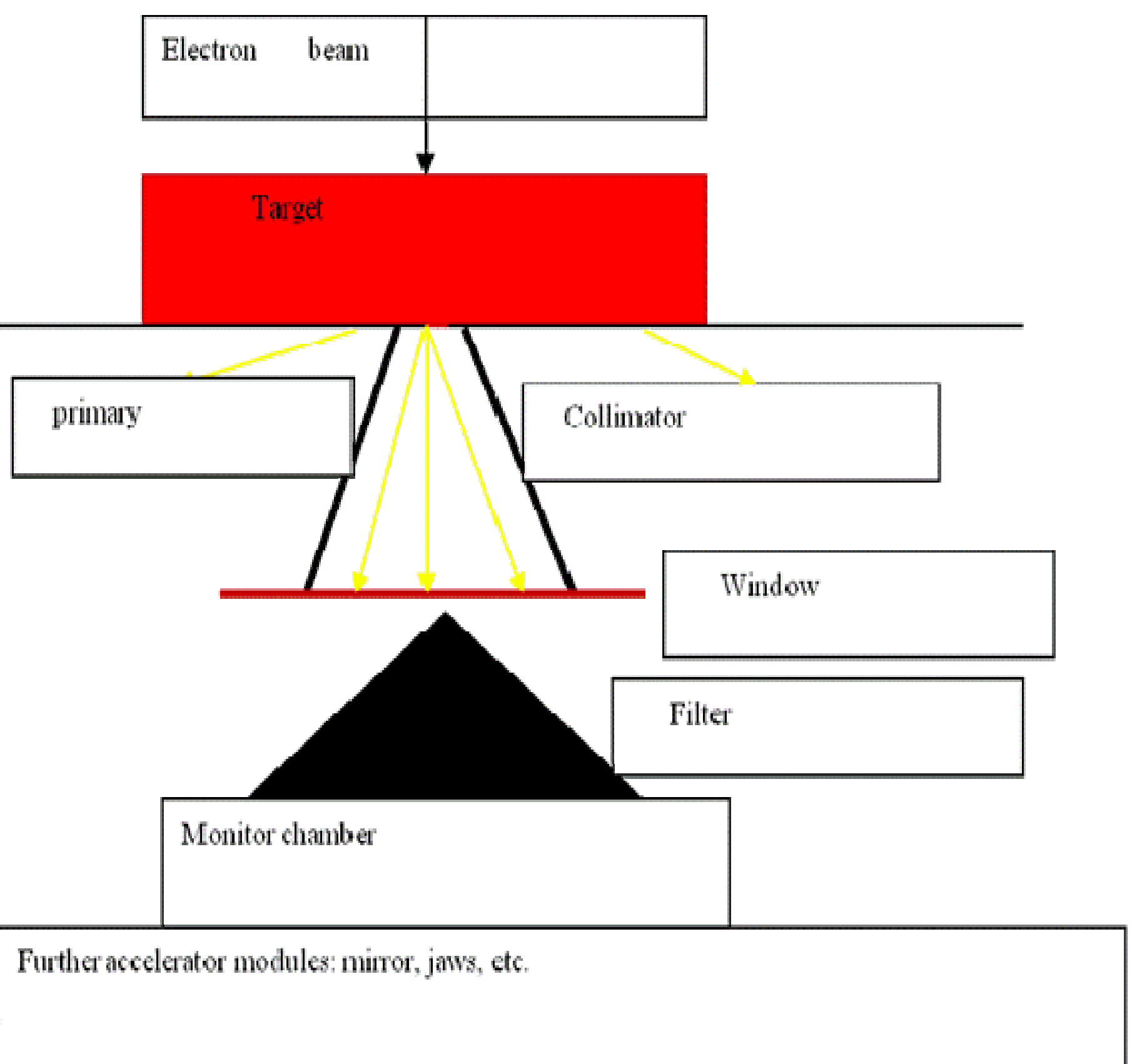

**Figure 1:** Schematic representation of a linear accelerator.

## 2.2. Qualitative considerations with regard to focusing of electron scatter

Figures 2 and 3 indicate that by restriction to a single Tungsten target there is no mean to prevent the scatter of electrons within the target material.

The thickness of the target of standard linear accelerators amounts to ca. 1 mm Tungsten and immediately below 1 mm Copper in order to increase the removal of the produced heat from the target. The bremsstrahlung spectrum created in Copper is significantly lower than that of Tungsten; it is most widely absorbed in the flattening filter. Since the multiple scatter and heat production are responsible for the low efficiency and for a lot of necessary shielding of the accelerator head, we consider at first an alternative way to exploit bremsstrahlung by a multitarget. This consists of a configuration of very thin Tungsten layers. There thickness should not exceed 0.04 mm. We assumed previously a thickness of 0.01 mm for each plate we need 100 plates to reach an overall thickness of 1 mm with an effective depth of 10 cm. However, with regard to clinical accelerators it is possible to work with a smaller number of plates. Figure 3 qualitatively shows that a tilted wall makes the reflection angle small and therefore the backscatter can be increased. The left-hand side of this figure presents the consequence of this effect, namely a cone configuration of the wall, which embeds 31 Tungsten plates. The heat production in each plate is negligible and no additional removal of heat is needed. Thus we can verify that the configuration below makes the primary collimator itself to a cone target consisting of ca. 100 plates and the created bremsstrahlung shows the preference direction of a cone. This configuration
5 implies two advantages: The total depth of cone is assumed to amount to 100 mm, and the 100 plates with 1 mm distance between the plates can be regarded as a continuum, i.e., a Tungsten density $\rho_t$ of the multitarget part of the cone can be assumed. Therefore theoretical calculations can be simplified much. Otherwise, we have to perform complicated numerical step-by-step calculations (this is only possible with regard to Monte-Carlo calculations). The second advantage refers to the direction of the
10 created bremsstrahlung. If the electron energy is much higher than 0.511 MeV (rest energy of the electrons), then the direction of the created bremsstrahlung approximately agrees with the direction of the impinging electrons. Thus bremsstrahlung with large angles cannot be completely avoided, but significantly reduced to a minimum contribution. In a second order, there exists also a focusing effect of bremsstrahlung at the Tungsten wall due to the small-angle part of the Compton effect.

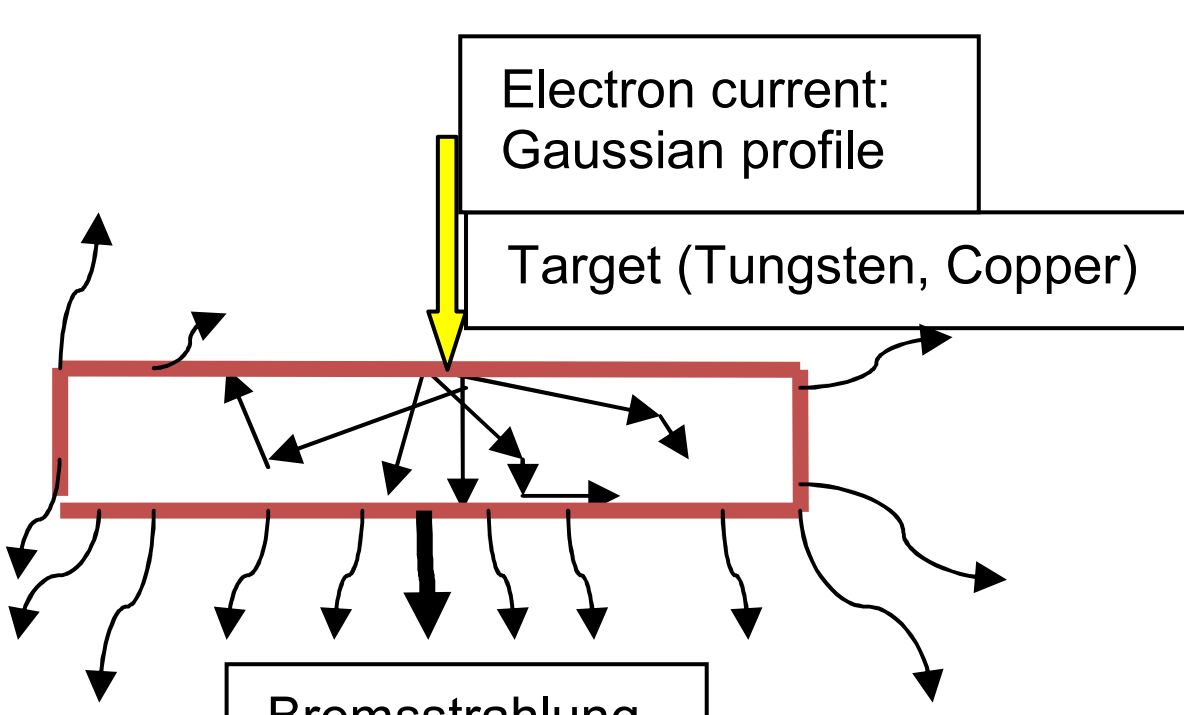

**Figure 2:** Multiple scatter and creation of bremsstrahlung in the target.

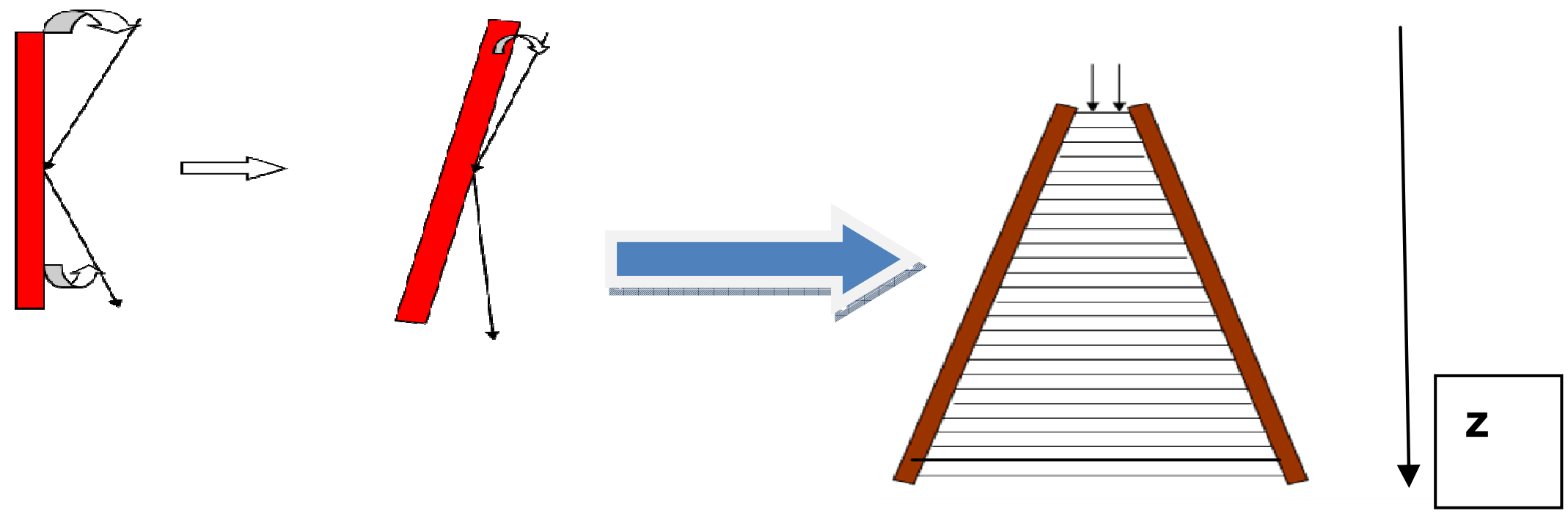

**Figure 3:** Small angle backscatter of electrons at a Tungsten wall (reflection) induces a focusing of electrons. The backscatter is amplified by a cone configuration of the Tungsten wall.

According to Figure 4 we can exploit and, by that, amplify the focusing influence obtained by wall
5 scatter, namely by an additional external magnetic field, which must have the property that a permanent gradient of the field component $B_r$ perpendicular to the propagation axis (z-axis) is present. Thus the complete configuration represents a similarity to features of an electron microscope. The effect of the additional magnetic field is:

1. Reduction of the reflection angle at the Tungsten wall.

10 2. Reduction of the impinging angle of the inner electrons at each Tungsten plate.

The additional magnetic field makes sense, if the cone is rather narrow, i.e. the opening angle is small and a very narrow radiation beam with extremely high intensity is required. However, the application of an additional magnetic field for focusing is optional, since many conceptual designs of medical accelerators can already be improved by a configuration without external magnetic field (Figure 3). It
15 should be pointed out that in both Figures 3 and 4 we have used a qualitative presentation based on forestalled results of succeeding sections. The new conceptions of designing linear accelerators can be justified by the following synopsis:

For electrons with energy >> $mc^2$ (0.511 MeV) the direction of the created γ-quanta agrees with the actual direction of motion of electrons. This implies for electrons, which suffered large-angle
20 scattering, that the created bremsstrahlung has unfortunately not the preference direction of the incident beam. There is additionally a non-negligible amount of energy loss of electrons and heat production as a consequence of multiple electron scatter.

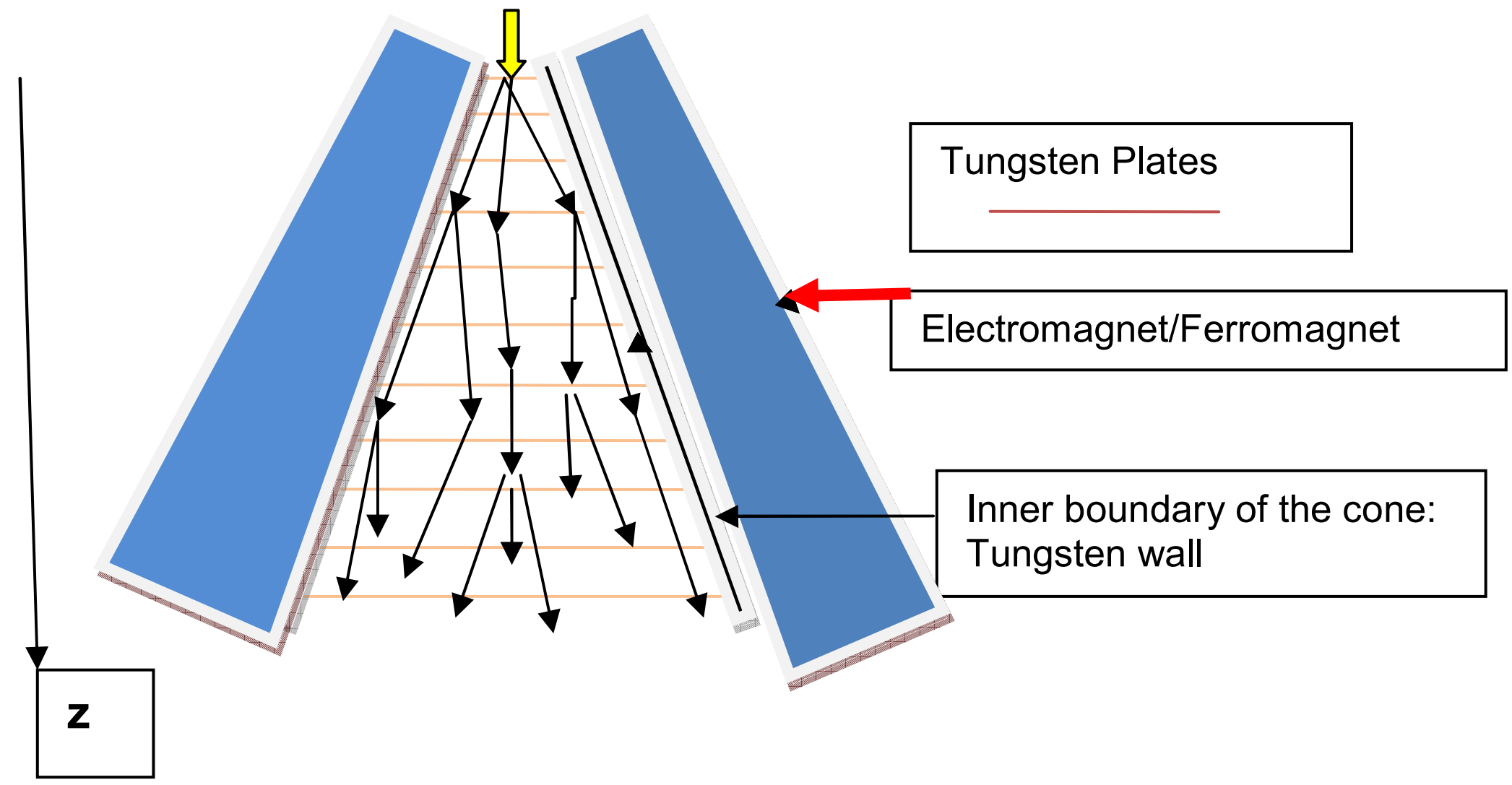

**Figure 4:** Schematic representation of a multi-layered target with an additional magnetic field for focusing.

A multi-layer target is suitable to reduce multiple scatter significantly by exploiting focusing effects by the cone wall (high Z material, e.g. Tungsten) and optionally by a proper magnetic field. Thus we have to split the conventional Tungsten target (thickness: usually 1 mm) in, at least, 10 subtargets whose thickness is of the order 0.1 mm and the distance between the layers should then amount to ca. 1 cm. However, a thickness of 0.012 mm (these layers can be purchased) and a distance of ca. 1 - 2 mm between each appears to be much more convenient.

The geometrical configurations of the three cases under consideration are:

*Case 1 (magnetic field necessary)*

Diameter of the Tungsten plate at entrance: 3 mm, at the end of the cone with z = 3 cm: 5 mm; diameter of the circular field size at z = 90 cm: ca. 6 cm.

*Case 2 (magnetic field optionally possible)*

Diameter of the Tungsten plate at entrance: 3 mm, at the end of the cone with z = 3 cm: 7 mm; diameter of the circular field size at z = 90 cm: ca. 12 cm.

*Case 3 (magnetic field not required)*

Diameter of the Tungsten plate at entrance: 3 mm, at the end of the cone with z = 3 cm: 13 mm; diameter of the circular field size at z = 90 cm: ca. 30 cm.

## 2.3. Theoretical calculations and Monte-Carlo calculations with GEANT4

### *2.3.1. Some physical toolkits*

*2.3.1.1. General properties and requirements*

With respect to both calculation procedures (GEANT4 and analytical calculations) we have to use the relativistic energy-momentum relation:

5 $$W^2 = m^2c^4 + \vec{p}^2 \cdot c^2 \quad (1)$$

The relativistic energy E of a particle (without rest energy) is given by:

$$\left.\begin{aligned} E &= mc^2/\sqrt{1-\beta^2} - mc^2 \\ \beta &= v/c \end{aligned}\right\} \quad (2)$$

The relativistic energy-momentum relation in the presence of a magnetic field (vector potential **A**) reads:

10 $$W^2 = m^2c^4 + (\vec{p} - (e/c)\cdot\vec{A})^2 \cdot c^2 \quad (3)$$

Equation (3) represents a quantum-mechanical equation, if the transition $\vec{p} \rightarrow \frac{\hbar}{i}\nabla$ is carried out. The components of the magnetic induction **B** with div·**B** = 0 are given by:

$$\left.\begin{aligned} B_x &= \partial A_y/\partial z - \partial A_z/\partial y \\ B_y &= \partial A_z/\partial x - \partial A_x/\partial z \\ B_z &= \partial A_x/\partial y - \partial A_y/\partial x \\ B_r^2 &= B_x^2 + B_y^2 \end{aligned}\right\} \quad (4)$$

The geometrical configuration is shown in Figure 5.

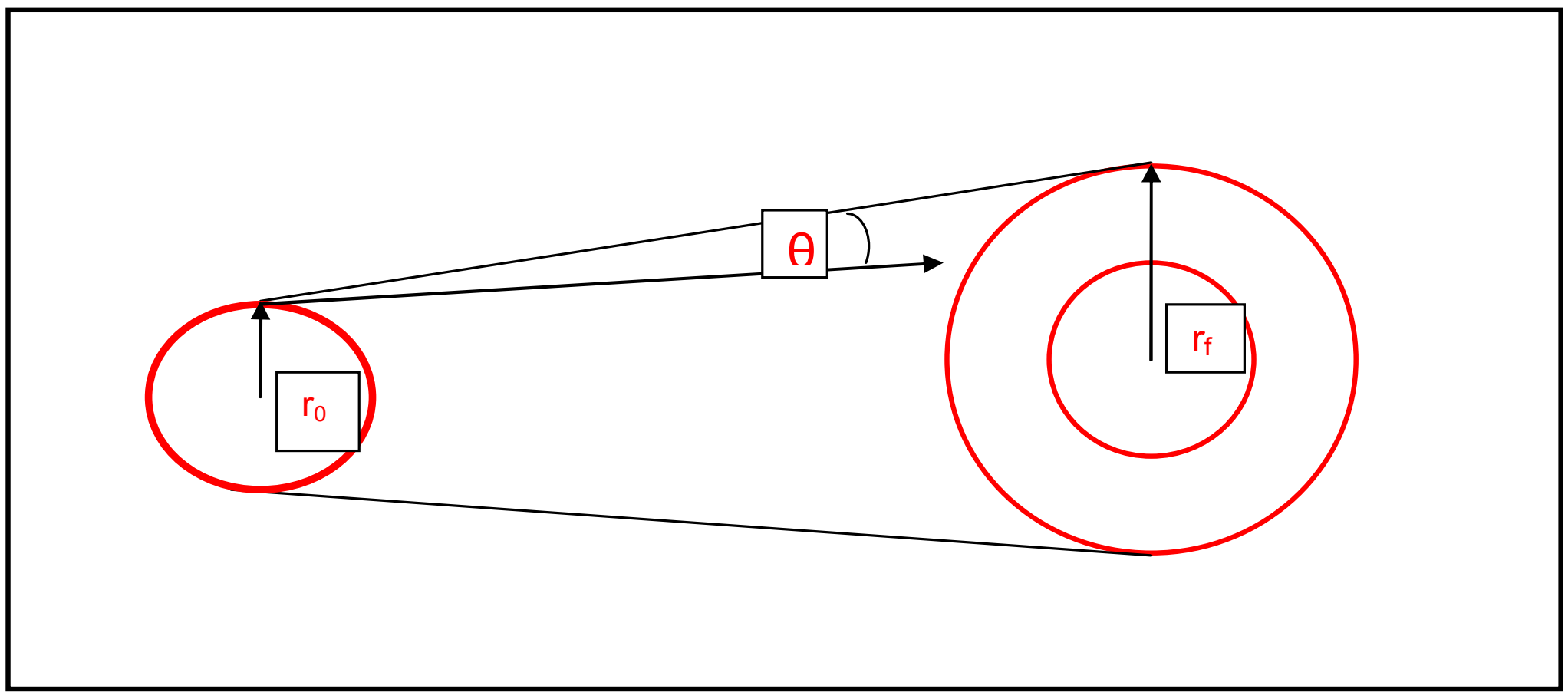

**Figure 5:** Schematic representation of the geometrical properties. On the left-hand side: Small circle of the cone entrance (radius $r_0$) for impinging electron beam. Right-hand side: Larger circle at the cone exit (radius $r_f$). The inner circle symbolizes the area of the entrance. The angle **θ** shows the opening angle of the cone, which is usually small in those cases, where a strong magnetic field is required. This will be verified in the result section.

20 The field strength has to satisfy, at least, the following properties:

$$\left.\begin{aligned} B_r(z) &= r^2 \cdot B_0 / r_0^2 + (z/L) \cdot (B_f - B_0) \cdot r^2 / r_f^2 \\ r &= r_0 + z \cdot \tan\vartheta = \sqrt{x^2 + y^2} \\ r_f &= r_0 + L \cdot \tan\vartheta \end{aligned}\right\} \quad (5)$$

For the reason of symmetry the following condition has to be valid:

$$|B_x| = |B_y| \quad (6)$$

In the above equation $B_r(z)$ means the radial component as a function of z, $r_0$ is the field radius at the entrance of the beam (z = 0), and $r_f$ the related radius at the end (z = L). The possible devolution of some cases of interest is shown in Figure 11. However, the properties of this figure do not represent a rigid scheme, since it mainly depends on the design of the target, whether other lengths L of the cone (central axis) are required. We want also point out that the magnetic field does not disappear at z = L. This length has only to agree with the focusing part of the magnetic field, whereas a small area with a defocusing part cannot be avoided. It may serve to remove those electrons with sufficiently low energy, where the production of bremsstrahlung is no longer desired.

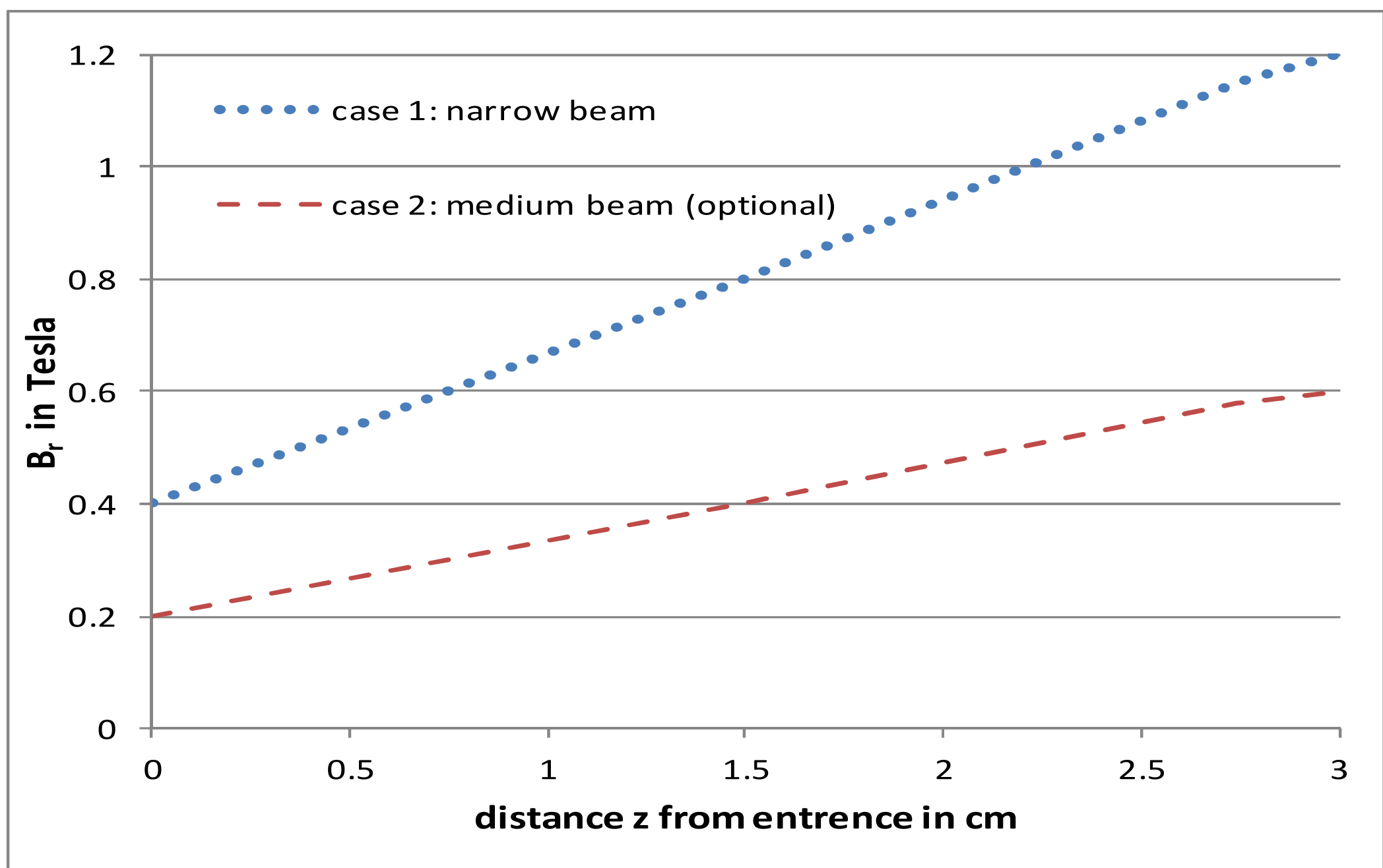

**Figure 6:** Increase of the radial component $B_r$ along the surface of the cone. $M_1$ = case 1: $B_0$ = 0.4 Tesla, $B_f$ = 1.2 Tesla; case 2: $B_0$ = 0.2 Tesla, $B_f$ = 0.6 Tesla.

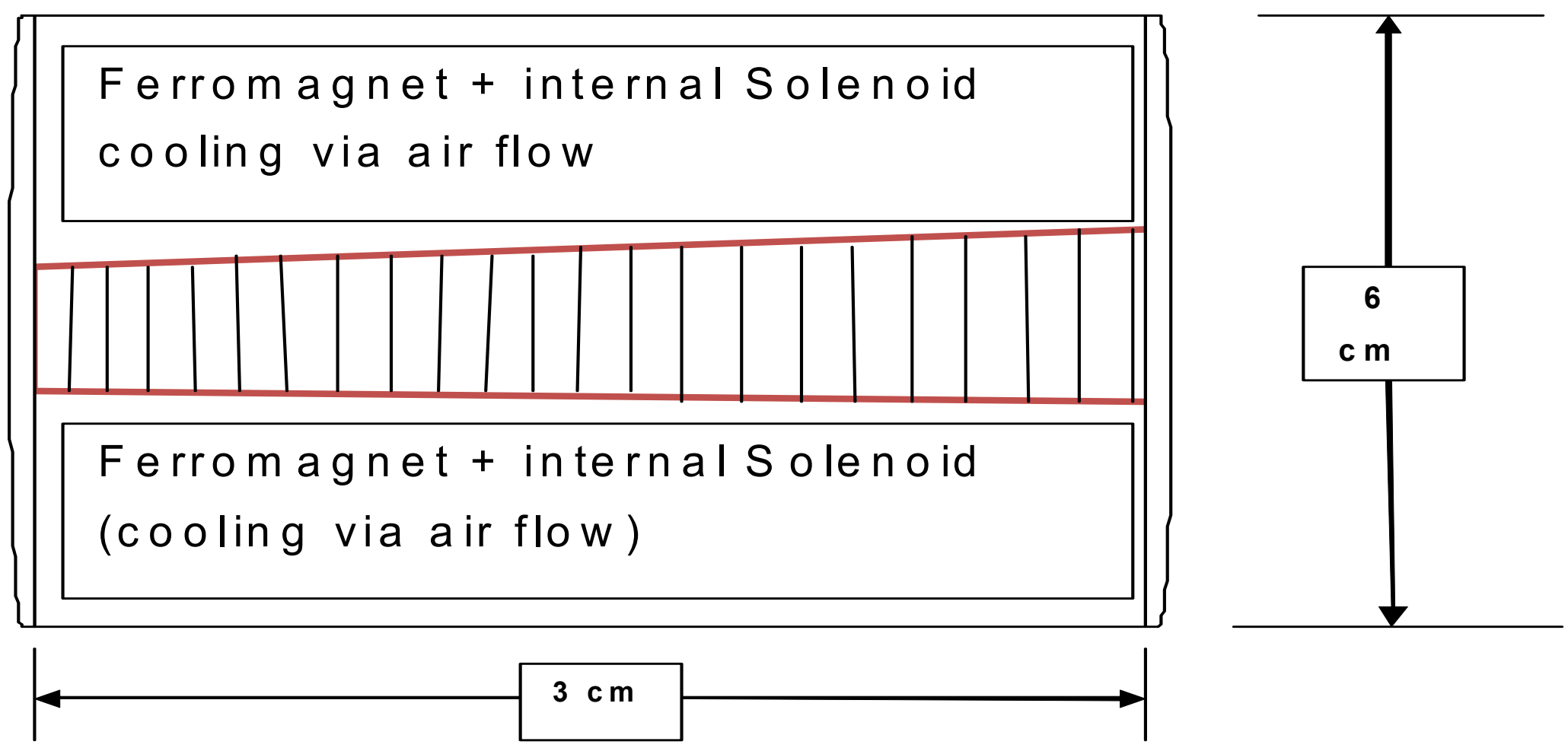

**Figure 6a:** Schematic representation of the considered configuration.

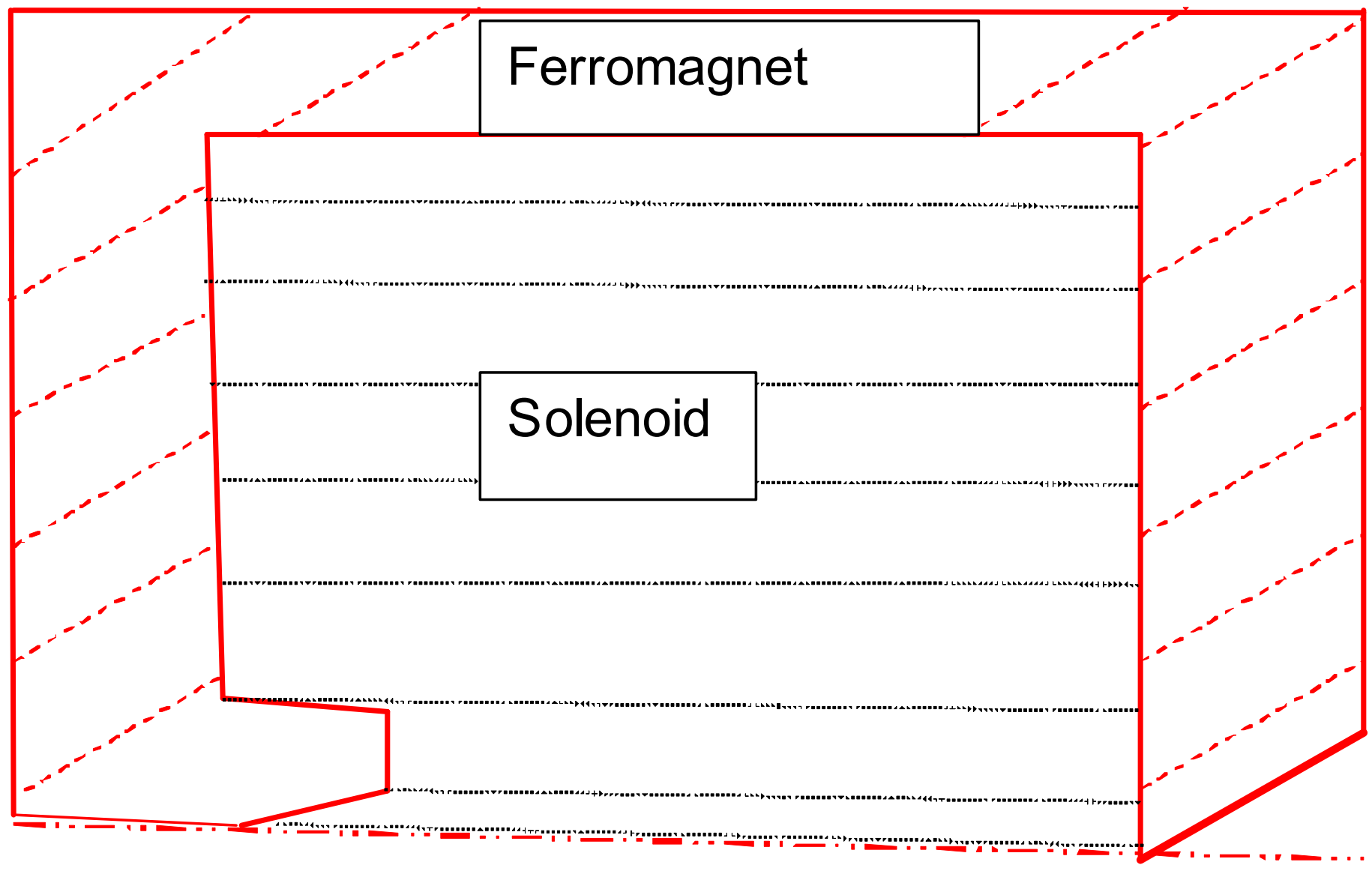

**Figure 6b:** Section of the magnet with rotational symmetry.

5 *2.3.2. Fermi-Eyges theory, multiple scatter theory of Molière and inclusion of magnetic fields*

Fick's law of diffusion plays a key role in a lot of physical/chemical/physiological processes, it is also used for the description of scatter and absorption of electrons, protons or neutrons in a medium such as Tungsten; it is referred to as Fermi-Eyges age equation (Eyges, 1948). A more accurate theory of scatter has been given by Molière (Molière 1955)

$$-\frac{\partial}{\partial t}\rho + D\,\Delta\,\rho = 0 \quad (7)$$

With respect to Fick's law D is the diffusion coefficient, ρ the particle (electron) density (concentration) and Δ the Laplace operator. In the case of the Fermi-Eyges theory, the particle density ρ has to be replaced by an energy-distribution density E, which reads:

$$-\frac{\partial}{\partial t}E + D_F\,\Delta\,E = 0 \quad (8)$$

Since both equations formally agree, we now introduce the amplitude ***U*** in order to be independent of the actual meaning. The same fact is also valid with regard to the constant factor D, which may either be identified with a diffusion constant D or with a parameter $D_F$ in Fermi-Eyges theory. Equations (7 - 8) follow from the property that a gradient of the density is connected with a current:

$$j = -D \cdot \nabla U \quad (9)$$

In addition, a balance equation has also to hold:

$$\left.\begin{aligned} \frac{\partial}{\partial t}U + div j &= 0 \\ div\; j &= (\nabla v)U \end{aligned}\right\} \quad (10)$$

The term $\nabla \cdot v$ represents the scalar product of the 3D differential operator $\nabla$ with a 3D velocity **v**. If the particles have the charge q (for electrons: q = e) and a magnetic field, described by the vector potential **A**, is present, then an additional momentum/velocity due to the magnetic field (Lorentz force) has to be accounted for:

$$\left.\begin{aligned} p &= -\frac{e}{c}A \\ v &= p/m = -\frac{e}{c\,m}A \\ divA &= \nabla \cdot A = 0 \end{aligned}\right\} \quad (11)$$

The velocity **v** results from the division of the momentum **p** by the particle mass M (= electron mass m) and c is the velocity of light. Now equations (9 - 10) read:

$$\left.\begin{array}{l} \boldsymbol{j} = -\left(\boldsymbol{D}\cdot\nabla - \frac{e}{mc}\boldsymbol{A}\right)\boldsymbol{U} \\ \frac{\partial}{\partial t}\boldsymbol{U} + div\boldsymbol{j} = 0 \\ -\frac{\partial}{\partial t}\boldsymbol{U} + \boldsymbol{D}\cdot\Delta\boldsymbol{U} - \frac{e}{mc}(\boldsymbol{A}\cdot\nabla)\boldsymbol{U} = 0 \end{array}\right\} \quad (12)$$

The last equation can also be written in the form

$$-\frac{\partial}{\partial t}\boldsymbol{U} + \boldsymbol{D}\cdot\Delta\boldsymbol{U} - div\,(\boldsymbol{v}\,\boldsymbol{U}) = 0 \quad (13)$$

If we replace U by a probability distribution P, then equation (13) turns out to be the Kolmogorov forward equation, and we regard $v$ as an effect of a magnetic field. However, equations (12 -13) are not yet fully gauge invariant, since we have only modified the diffusion current **j** by the magnetic interaction. The magnetic effect is not yet included in the balance equation. In order to complete this condition we have to write:

$$\left.\begin{array}{l} \frac{\partial}{\partial t}\boldsymbol{U} + \left(\nabla - \frac{e}{mcD}\boldsymbol{A}\right)\left(\boldsymbol{D}\cdot\nabla - \frac{e}{mc}\boldsymbol{A}\right)\boldsymbol{U} = 0 \\ -\frac{\partial}{\partial t}\boldsymbol{U} + \boldsymbol{D}\cdot\Delta\boldsymbol{U} - \frac{2e}{mc}(\boldsymbol{A}\cdot\nabla)\boldsymbol{U} + \left(e^2\boldsymbol{A}^2 / \boldsymbol{D}\,m^2c^2\right)\boldsymbol{U} = 0 \end{array}\right\} \quad (14)$$

Besides gauge invariance a further reason for obtaining equation (14) is the transition to the Schrödinger equation with magnetic field, since the simple Fick's law of diffusion can be transformed to a Schrödinger equation without external fields. Equation (14) provides a correspondence in the presence of an external magnetic field. It should be noted that the Kolmogorov forward equation is contained as in the special case of Equation (26) by setting $\mathbf{A}^2 = \mathbf{0}$. According to the Fermi-Eyges/diffusion theory we can always put

$$\sigma^2 = 2Dt \;\; or \;\; \sigma^2 = 2D\tau \quad (15)$$

The parameter $\tau$ can replace the arbitrary time-variable t, if scatter/diffusion only occurs in a short time interval. We consider the case, where the z-component $B = B_z = B_0$ of a constant magnetic field is responsible for the motion of electrons and pass to the general case thereafter. Since the magnetic field strength $B_0$ is given by $\boldsymbol{B} = rot\,\boldsymbol{A} = \nabla x \boldsymbol{A}$ , we can choose **A** as follows:

$$\left.\begin{aligned} A_x &= -By/2,\; A_y = Bx/2 \\ B_z &= \partial A_y/\partial x - \partial A_x/\partial y = B_0 \\ A^2 &= B_0^2 (x^2 + y^2)/4 \end{aligned}\right\} \quad (16)$$

By that, Equation (27) becomes (U = E):

$$\left.\begin{aligned} &-\partial E/\partial t + D_F \,\Delta E + (e^2 B_0^2/4m^2c^2 D_F)(x^2+y^2)E \\ &-(eB/cm)[y\partial/\partial x - x\partial/\partial y]E = \lambda_1 E \end{aligned}\right\} \quad (17)$$

Since in z-direction motion without magnetic interaction is allowed, the following separation of the above equation is possible:

$$E = \varphi(x,y,t)\exp(-\lambda_1 t)\exp(ikz)\exp(-k^2 D_F t) \quad (18)$$

This provides the following equation:

$$\left.\begin{aligned} &(\partial^2/\partial x^2 + \partial^2/\partial y^2)\varphi - (\omega_0/D_F)[y\partial/\partial x - x\partial/\partial y]\varphi \\ &+(\omega_0^2/4D_F^2)(x^2+y^2)\varphi = (1/D_F)\,\partial\varphi/\partial t \end{aligned}\right\} \quad (19)$$

The following parameters are given by:

$$\omega_0 = eB_0/mc \quad (20)$$

$$\lambda = \lambda_1 + \lambda' \quad (20a)$$

It has to be pointed out that the Larmor frequency $\omega_0$ according to equation (20) is identical with those obtained by a Schrödinger or Dirac equation with an external magnetic field. The basis solution (generating function) from which we can construct all other solutions (see e.g. comparison with Schrödinger equation) is given by

$$\varphi(x,y,t) = \exp[-a(x^2+y^2)/2]\exp(-\lambda' t) \quad (21)$$

$$\lambda' = \pm i\,\omega_0 \quad (22)$$

$$a^2 = -\omega_0^2/4D_F^2 \Rightarrow a = \pm i\omega_0/2D_F \quad (22a)$$

The parameters *a* and *λ'* represent complex values; however, we can form a linear combination to get the real solution: Before we shall consider this property, we note that with respect to the z-coordinate a set of solutions of the diffusion equation are permitted (see Ulmer 1983). They result from the Fourier transform of the specified z-dependent function:

$$\left.\begin{aligned}&\Phi(z,t)=\frac{1}{\sqrt{2\pi}}\int_{-\infty}^{\infty}A(k)\exp(ikz)\exp(-k^2D_F\,t)dk\\&A(k)=a_0+a_1k+a_2k^2+...+a_nk^n\end{aligned}\right|\quad(23)$$

Every power of k of the expansion of A(k) represents a solution, if the integral is carried out. We therefore denote

$$\Phi_n(z,t)=P_n(z,t)\exp(-z^2/4D_Ft)\quad(24)$$

The polynomials $P_n(z, t)$ have been previously defined by evaluation of the above integral (23):

$$\Phi(z,t)=\frac{1}{\sqrt{2\pi}}\int_{-k_0}^{k_0}\exp(ikz)\exp(-Dk^2t)dk/2k_0\quad(25)$$

The boundary parameter $k_0$ is given by $k_0 = 1/z_0$. The evaluation of the integral (18) provides

$$\left.\begin{aligned}&\Phi(z,t)=U(t)\exp(-z^2/4D_Ft)[erf(s_1)-erf(s_2)]\\&U(t)=z_0/4\sqrt{2D_Ft}\\&s_1=(\;1/z_0+iz/2D_Ft)/\sqrt{D_Ft}\\&s_2=(-1/z_0+iz/2D_Ft)/\sqrt{D_Ft}\end{aligned}\right\}\quad(26)$$

With the help of equation (26) we consider the solution:

$$\left.\begin{aligned}&E(x,y,z,t)=\Phi(z,t)\cdot\exp(-\lambda_1t)\varphi\}\\&\varphi=A_0\cos(\omega_0(x^2+y^2)/4D+\omega_0t)+A_1\sin(\omega_0(x^2+y^2)/4D_F+\omega_0t)\end{aligned}\right\}\quad(27)$$

The sine and cosine function appears by forming linear combinations of solutions of equation (21), since a according to equation (23) is an imaginary parameter and the theorem for complex exponential functions can be applied. It has to be mentioned that the cosine as well as the sine are solutions, and both may form linear combinations according to equation (26).We should account for that the function $\varphi$ in equation (18) has not to be restricted to the simple sine and cosine, but we can also use the general solution manifold according to equations (26 - 27).

Before we shall study some properties of equation (27), a comparison with the Schrödinger equation is indicated.

The free particle Schrödinger equation is given by

$$i\hbar\,\partial\Psi/\partial t=-(\hbar^2/2m)\Delta\Psi\quad(28)$$

It assumes the character of an irreversible transport equation, if the substitution $t = i\tau$ is carried out. By that, the diffusion constant is given in terms of the Planck's constant: $D_F = \hbar/2m$.

However, the solution (27) is not the only possible one, and we are able to obtain a spectrum of solutions and their linear combinations. The complete solution spectrum is given by the two different types:

Powers of even order:

$$\left.\begin{aligned}
\varphi_{2n} = \sum_{m=0}^{n} \{ A_{2m} \cos^{2m}((\omega_{2n}/4D_F)(x^2+y^2)+\omega_{2n} t) \\
+ \quad B_{2m} \sin^{2m}((\omega_{2n}/4D_F)(x^2+y^2)+\omega_{2n} t) \\
\omega_{2n} = \omega_0 \cdot 2n \;\; (n \geq 1, n = 1,2,3,)
\end{aligned}\right\} \quad (29)$$

Powers of odd order:

$$\left.\begin{aligned}
\varphi_{2n+1} = \sum_{m=0}^{n} \{ A_{2m+1} \cos^{2m+1}((\omega_{2n+1}/4D_F)(x^2+y^2)+\omega_{2n+1} t) \\
+ \quad B_{2m+1} \sin^{2m+1}((\omega_{2n+1}/4D_F)(x^2+y^2)+\omega_{2n+1} t) \\
\omega_{2n+1} = \omega_0 \cdot (2n+1) \;\; (n \geq 0, n = 0, 1, 2, 3,)
\end{aligned}\right\} \quad (30)$$

Please note that superpositions of different order and related eigen-frequencies are also possible solutions. Thus we can perform a linear combination of all solutions, e.g. a fast oscillating solution with a slow oscillating solution can be combined to form beat oscillations.

At first, we look at the connection between diffusion and the quantum mechanical Schrödinger equation with external magnetic fields. The following aspects should be emphasized: The resonance conditions for $\omega_0$ are in both cases identical, this appears to be rather noteworthy. In a formal sense, we have only to substitute the real time t by an imaginary time $\tau \Rightarrow it$, and the reversibility of the Schrödinger equation goes lost. This behavior is also known from the path integral formulation according to Feynman and Hibbs, which represents a further possible way to solve the complicated task of scatter and the role of magnetic fields by perturbation theory.

With respect to the eigenfrequency $\omega_0$ and its dependence on the related parameters e, m and B we are able to make the following statements:

The z-part of the solution has also the character of an oscillator due to the complex argument yielding nodes (see e.g. the book of Abramowitz and Stegun 1970). Only for sufficient large time $t \Rightarrow \infty$ a homogeneous charge distribution will be reached. The x – y – part does not allow broadening by diffusion. The behavior is comparable to that of a magnetic lense. Let us now consider an example of a magnetic bifurcation. Assume an oscillating propagation in the x-y plane with the highest frequency $\omega_0$ given by the magnetic field strength $B_0$. Thus a sudden change of the magnetic field strength from + $B_0$ to $B_0' = B_0 + \Delta B_0$ leads to a magnetic bifurcation, and, in particular, the antisymmetric sine functions change the sign, when the argument becomes negative. Such an effect may be induced by an inhomogeneous magnetic field yielding changes of the field strength (amount and orientation). The symmetry is spontaneously broken. The same fact may also happen under a lot of similar external influences: The change of the homogeneity of the magnetic field yields a change of the diffusion constant $D_F$; a change of the energy distribution E may require the formation of complete different patterns and oscillation frequencies, etc.

A principal result of the Bethe-Heitler theory is that the energy loss due to creation of bremsstrahlung is proportional to the actual electron energy. The differential equation for the radiation loss reads (in one dimension):

$$-dE_{bre} / dz = X_{rl}^{-1} \cdot E_{bre} \quad (31)$$

The theory of the creation of 'bremsstrahlung' can be formulated by the propagator method. The above mentioned phenomenological description summarizes all these parameters resulting from the quantum theoretical treatment by the radiation length $X_{rl}$ according to equation (31).

By iteration of equation (31) we obtain a second order differential equation, and the extension to 3D can readily carried out, i.e. the Laplace operator Δ appears. This extension has the advantage that the resulting equation can be added to further phenomenological equations containing the Laplace operator:

$$-\Delta E_{bre} = X_{rl}^{-2} \cdot E_{bre} \quad (32) .$$

A further advantage results from the previous Figure 3: If the amount of Tungsten sublayers is high, and, by that, the distance between them is small (e.g. 1 mm in the cone target), it is possible to solve equation (32) under continuum conditions. The total Tungsten mass can be divided by the cone volume to obtain the medium density $\rho_t$. Step-by-step calculations (we do not report them here) showed that for 2 mm distances between the plates and identical overall mass a continuum approximation can be justified.

In a phenomenological theory, we can summarize the complete problem by including both, energy loss by radiation loss (Bethe-Heitler theory) and energy dissipation (Fermi-Eyges theory)

$$\left.\begin{aligned} -\partial \mathbf{E}/\partial t &= \mathbf{D_F}\cdot\Delta\mathbf{E} - (2e/mc)\cdot(\vec{\mathbf{A}}\cdot\nabla)\cdot\mathbf{E} + \\ &(e^2\vec{\mathbf{A}}^2/\mathbf{D_F} m^2c^2)\mathbf{E} + \\ &\Delta\mathbf{E}_{bre} + \Delta\mathbf{E}_{col} = \mathbf{X}_{rl}^{-2}\cdot\mathbf{E}_{bre} + \mathbf{X}_{col}^{-2}\cdot\mathbf{E}_{col} \end{aligned}\right\} \quad (33)$$

In equation (33) the parameter $X_{rl}$ is referred to as the radiation length, which is proportional to $Z^2$, $N_A$ and $A_N$, whereas $X_{col}$ refers to the energy absorption by the cone wall (collimator), which is proportional to Z, $N_A$ and $A_N$. The nuclear charge is denoted by Z, the nuclear mass number by $A_N$, and $N_A$ is the Avogadro number.

The influence of the magnetic field can be accounted by the following solution expansion:

$$\left.\begin{aligned} E(x,y) &= \sum_{n=0}^{N}\{A_{2n}\cos((\omega_{2n}/4D_F)\cdot r^2) + \omega_{2n}\tau) + B_{2n+1}\cos((\omega_{2n+1}/4D_F)\cdot r^2) + \omega_{2n+1}\tau)\} \\ \omega_n &= neB'/mc = ne(B_0 + \Delta B_0)\ (\textit{Larmor frequency}) \end{aligned}\right\} \quad (34)$$

B' refers to as a correction of $B_0$ by $\Delta B_0$, since the magnetic induction must not be constant in the volume under consideration. In principle, we have to account for $N \rightarrow \infty$, which is impossible in numerical calculations.

**There are two possible procedures, which we have worked out:**

- **Solution of the scatter problem by a proper magnetic field acting between the subtargets and determination of the corresponding phase space for Monte-Carlo calculations (GEANT4) with respect to collision interaction (Bethe-Bloch) and bremsstrahlung (Bethe-Heitler theory).**
- **Complete solution of the above differential equation containing all 3 components using the tools given by deconvolution and inclusion of magnetic fields. In such a situation we have to put: $\mathbf{E_{bre} = E_{col} = E}$.**

The mathematical problem of scatter removal by deconvolution operators has been presented (Ulmer, 2010); the application with inclusion of magnetic fields for scatter removal has been analyzed previously (Ulmer, 2011).

2.3.3. *Monte-Carlo calculations with GEANT4*

GEANT4 (GEANT4- documents 2005) represents an open system of a Monte-Carlo code. Significant features with regard to our problem are creation of bremsstrahlung, multiple scatter according to Molière, heat production (Bethe-Bloch equation), energy straggling (Gaussian-Landau-Vavilov), Compton scatter of γ-radiation, and the actual energy/momentum of the electron after interactions leading to energy loss and change of the momentum. More sophisticated applications with regard to the focusing of a multi-layered Tungsten target and back scatter of the cone walls (Tungsten, Tantalum, Lead) under boundary conditions require the explicit use of the differential cross-section formula q(θ) with the form factor function F(θ). A further feature is the implementation of the magnetic field **B** (i.e. vector potential **A**) to account for the Lorentz force along the track of the electrons according to the relation (3).

*In order to obtain a reliable statistical foundation, each Monte-Carlo run has been performed with* $500 \cdot 10^6$ *histories.*

**Figure 7:** Back scatter properties (wall reflexion) of 9 MeV electrons at a high Z wall (W, Ta, Pb). The corresponding properties of 6, 18 or 20 MeV are rather similar.

Since the figures 7 – 10 have methodical character, we should like to show them already in this section. In particular, Figure 7 has a fundamental meaning in this study, namely angle-dependence of the reflexion (back scatter) of fast electrons at wall consisting of high Z material (Tungsten, Tantalum, and Lead). Although Pb shows the high Z value, the density is much smaller than that of W or Ta, and therefore according to Figure 7 we prefer Tungsten as the wall material for focusing. In particular, Figure 7 represents the essential properties used in Figure 3.

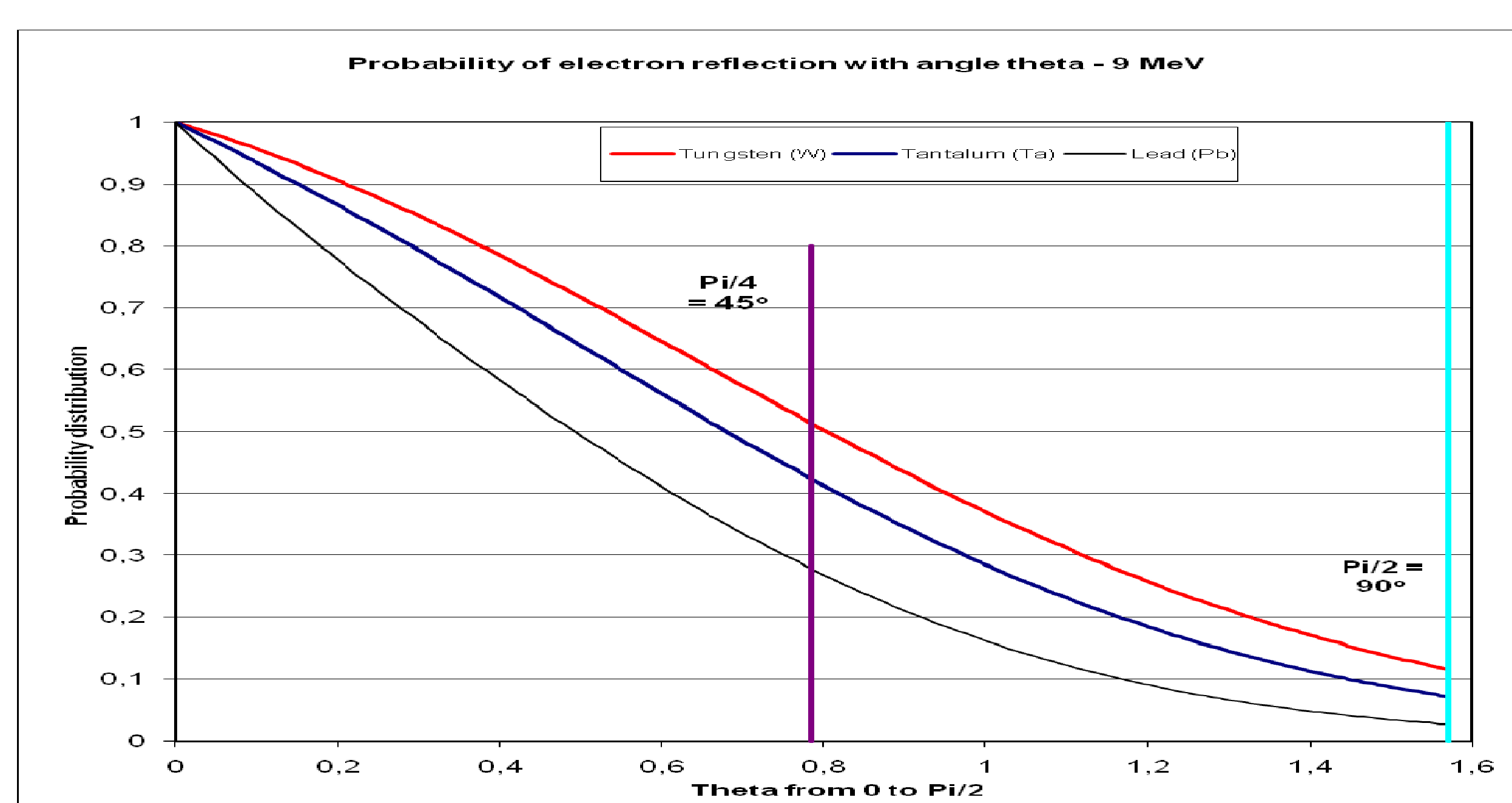

**Figure 7.** Backscatter (reflection) of fast electrons in dependence of the impinging angle.

Figure 8 shows the scatter behavior of fast electrons in air. In contrast to γ-radiation the scatter of electrons in air is not negligible. The initial condition in all 3 figures is an infinitesimally thin pencil ray of electrons. A consequence of these figures is that the multitarget has to be located in a vacuum in order to keep the lateral scatter of electrons as small as possible.

We should also point out that cross-section formulas used in the theoretical part of the foregoing publication (Ulmer 2010) would lead to a wrong behavior of reflection of electrons (e.g. for angles smaller than $20^0$), if the form factor function F(θ) would have been omitted, since q(θ) would then be highly diverging. On the other side, it is our particular interest to exploit small angle back scatter at the Tungsten wall.

At this place it should also be mentioned that a smaller focusing effect in the multilayer cone is obtained by the Compton scatter of the γ-radiation, if the γ-quanta are scattered inside the cone. However, the focusing of fast electrons is much more significant.

Since the focusing via wall scatter works best with Tungsten, we do, in general, not present calculations with other material such as Tantalum or Lead. The only exception with a Ta/Pb combination of the cone wall is restricted to one case in order to verify the preference of a Tungsten wall. In all figures of the section results we have adjusted the impinging electron beam to real conditions: The radial distribution at target surface is assumed to be a Gaussian with σ = 1 mm: $I(r) = I_0 \cdot exp(-r^2/(2 \cdot \sigma^2))$.

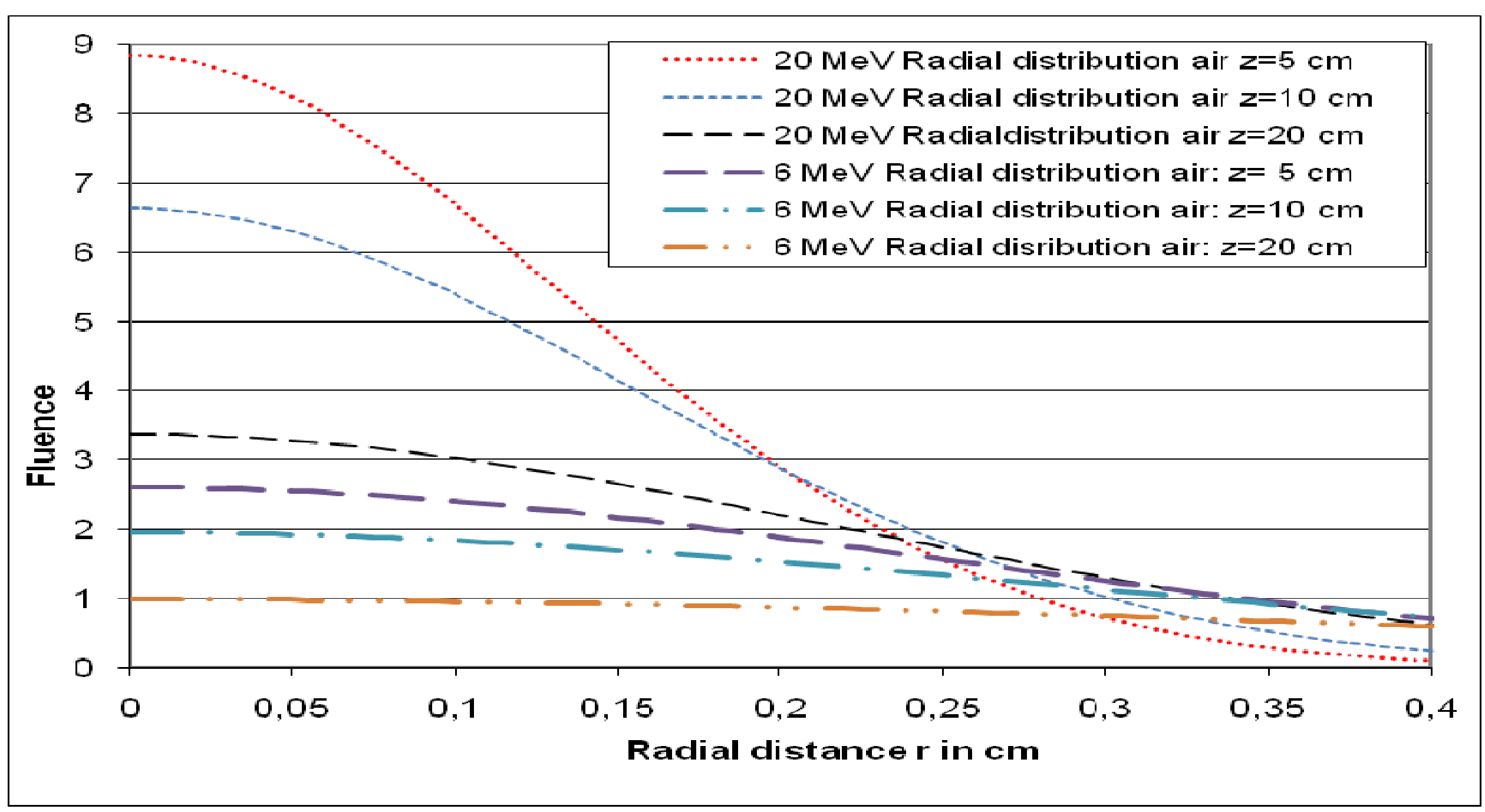

**Figure 8:** Comparison of air scatter of 6 MeV and 20 MeV electrons.

## 3. Results

The succeeding figure 11 serves as a reference standard for all other figures; this figure has been taken from the previous study (Ulmer 2011) and serves as a comparison standard. The bremsstrahlung production according to Figure 1 (blue curve, standard target) is scored along the plane immediately below the Tungsten target. The height at the central axis ($x = y = 0$) is normalized to *‘1’*, and the whole behavior of the intensity distribution shows all disadvantages of the conventional target, since it decreases slowly, and even at a radius of 7 cm a noteworthy intensity has been scored. Thus the domain with $r > 1$ cm results from multiple electron scatter in the target with no benefit for any application and requires a lot of shielding material. The behavior in the domain $r < 1$ cm gives raise to study a multitarget cone with a radius of 1 cm at the end of the cone. The cone consists of 20 layers (distance 5 mm per layer), total depth: 10 cm, the thickness of the wall amounts to 0.02 mm Tantalum (inside) and 10 mm Lead (outside) in contrast to all other cases, where 2 mm Tungsten have been used.

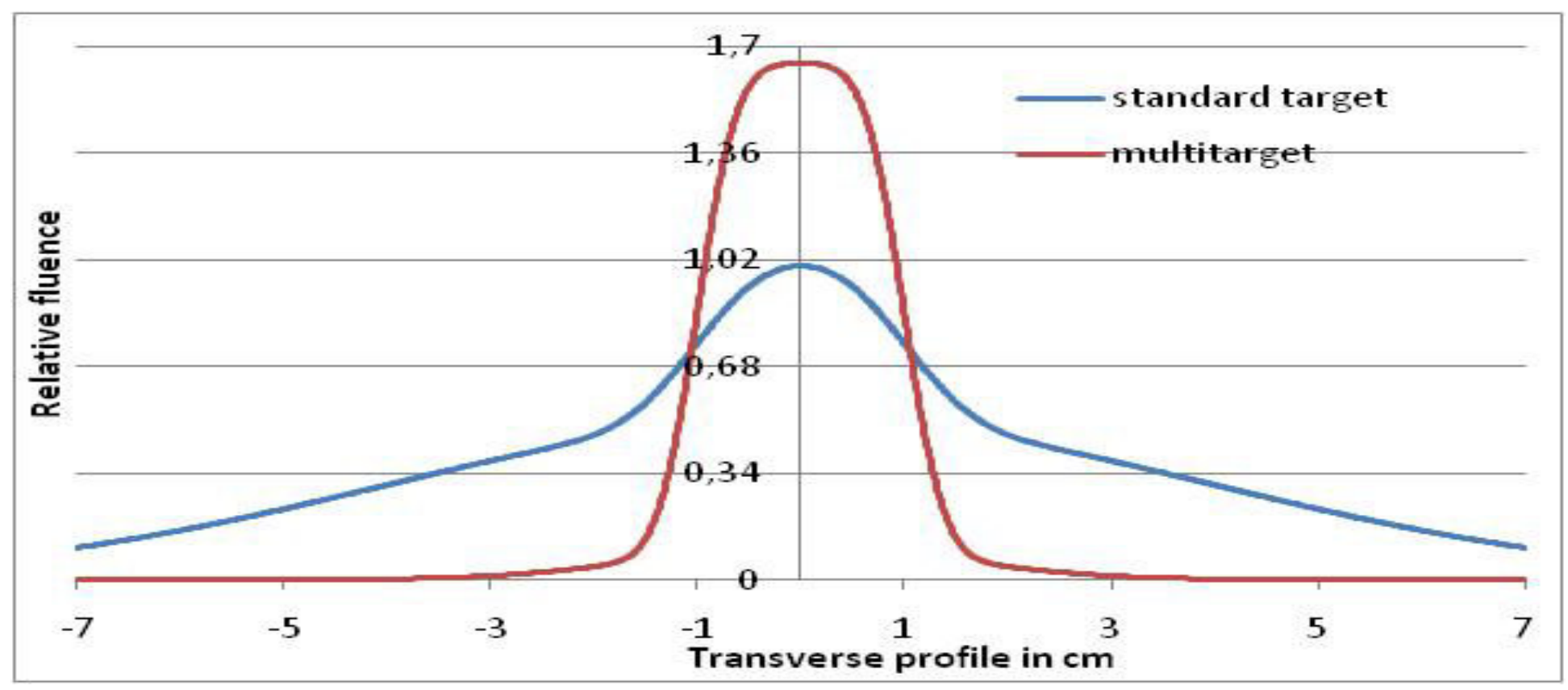

**Figure 11:** Comparison between standard target (Figure 1) and multi-layer target, electron energy E = 6 MeV.

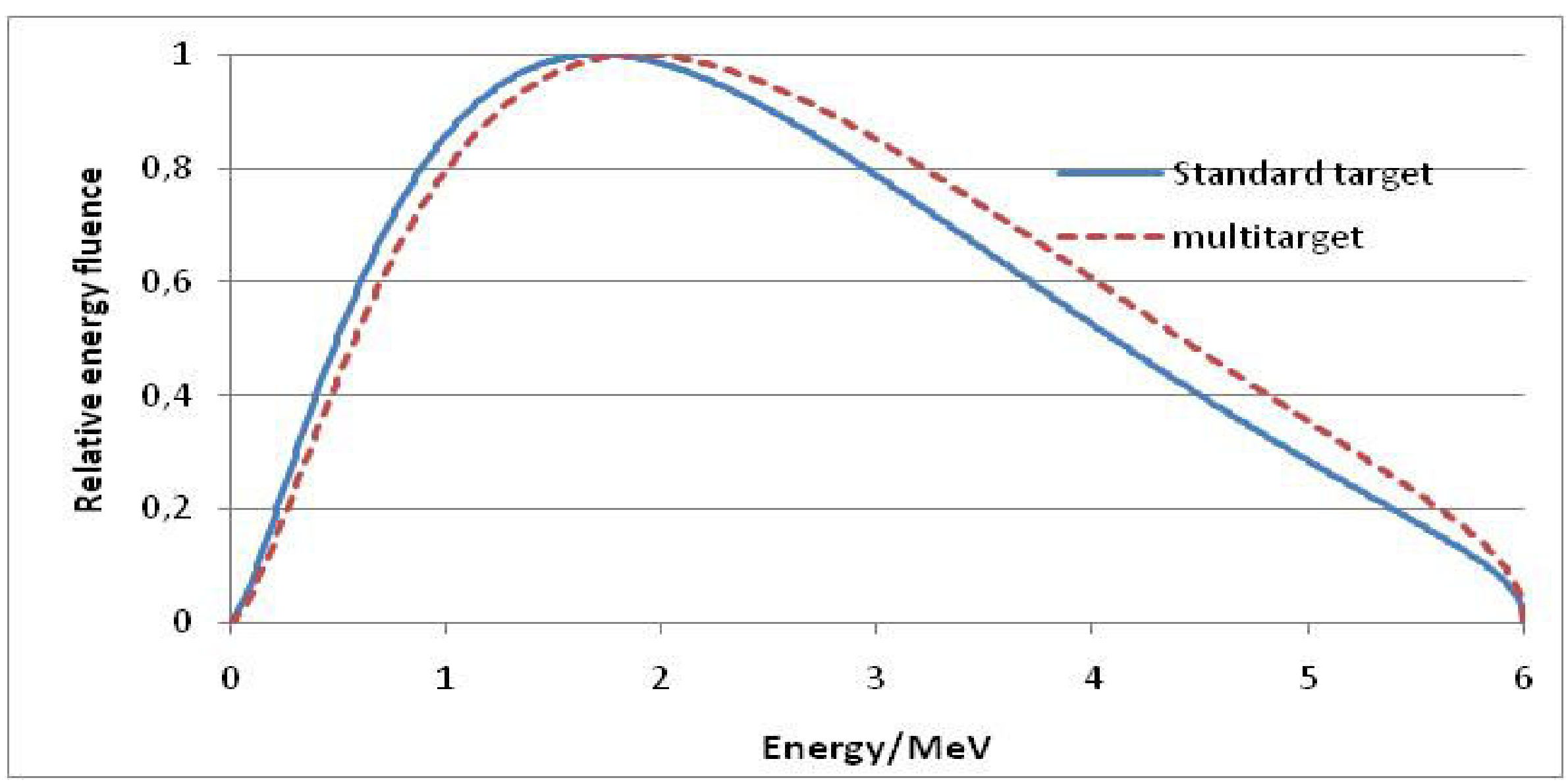

**Figure 12:** Relative energy fluence spectrum of the bremsstrahlung of 6 MeV electrons. The standard target refers to the condition presented by Ulmer et al (2005), i.e. below the flattening filter. In contrast to this condition the multitarget spectrum is scored at the end of the cone. The flattening filter is superfluous.

5 It should be pointed out that the application of E = 18 MeV electron energy instead of E = 6 MeV leads to rather similar properties as shown in Figure 11. Therefore we do not report them. With regard to all forthcoming Figures we use standard conditions of the cone wall, which consists of 2 mm Tungsten (with and without external magnetic field). It should be noted that in all results we had to assume air between the plates, the cones were not positioned in a vacuum.

10 Now we want to turn the interest to the three cases according to section 2. We should add that for comparison we have also considered the case, where the Tungsten wall has been replaced by Lucite. Figures 14 – 16 show calculated results immediately below the exit of the photon beam at the cone end, where measurements were impossible.

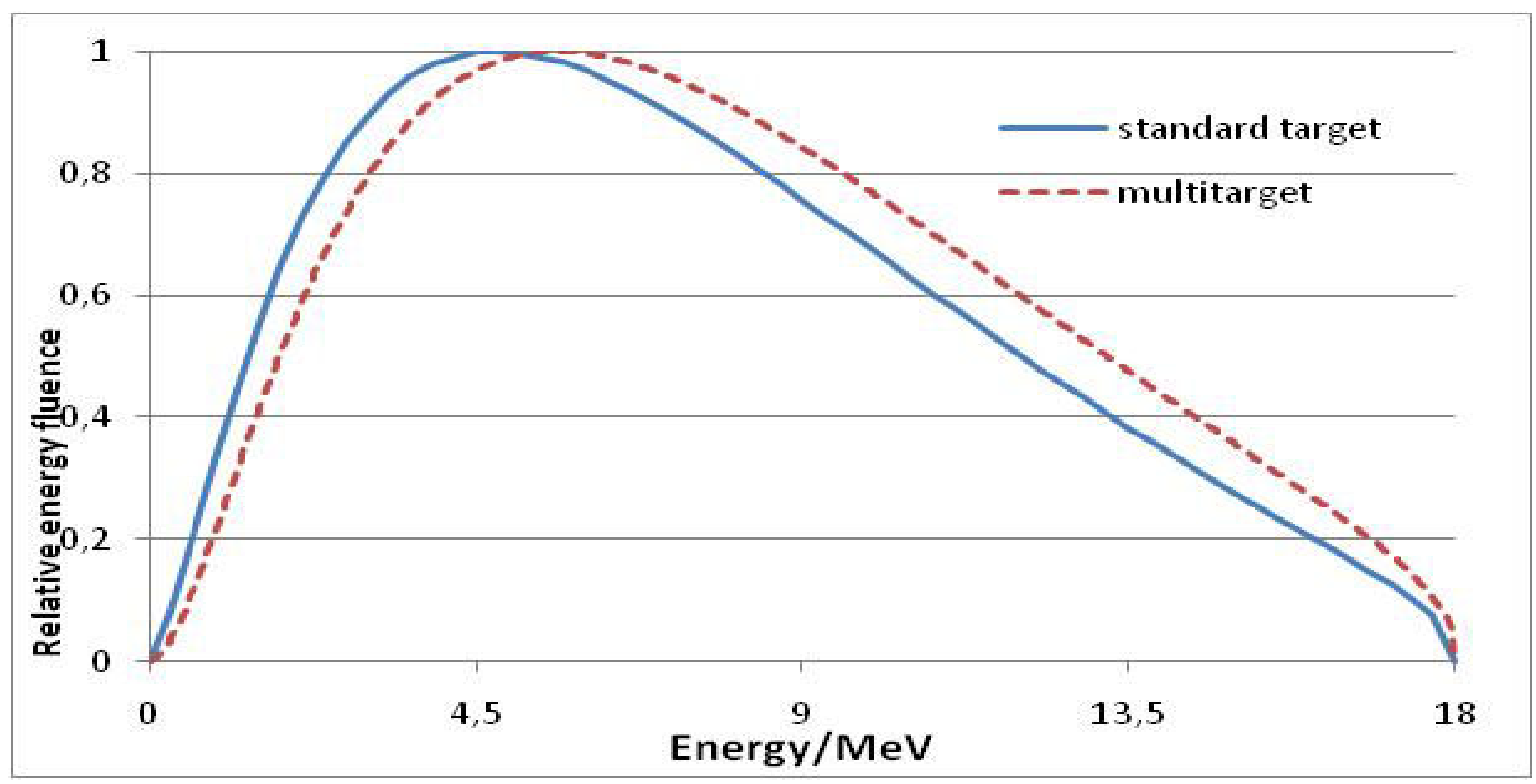

**Figure 13:** Relative energy fluence spectrum of the bremsstrahlung of 6 MeV electrons. The standard target refers to the condition presented by Ulmer et al (2005), i.e. below the flattening filter. In contrast to this condition the multitarget spectrum is scored at the end of the cone. The flattening filter is superfluous.

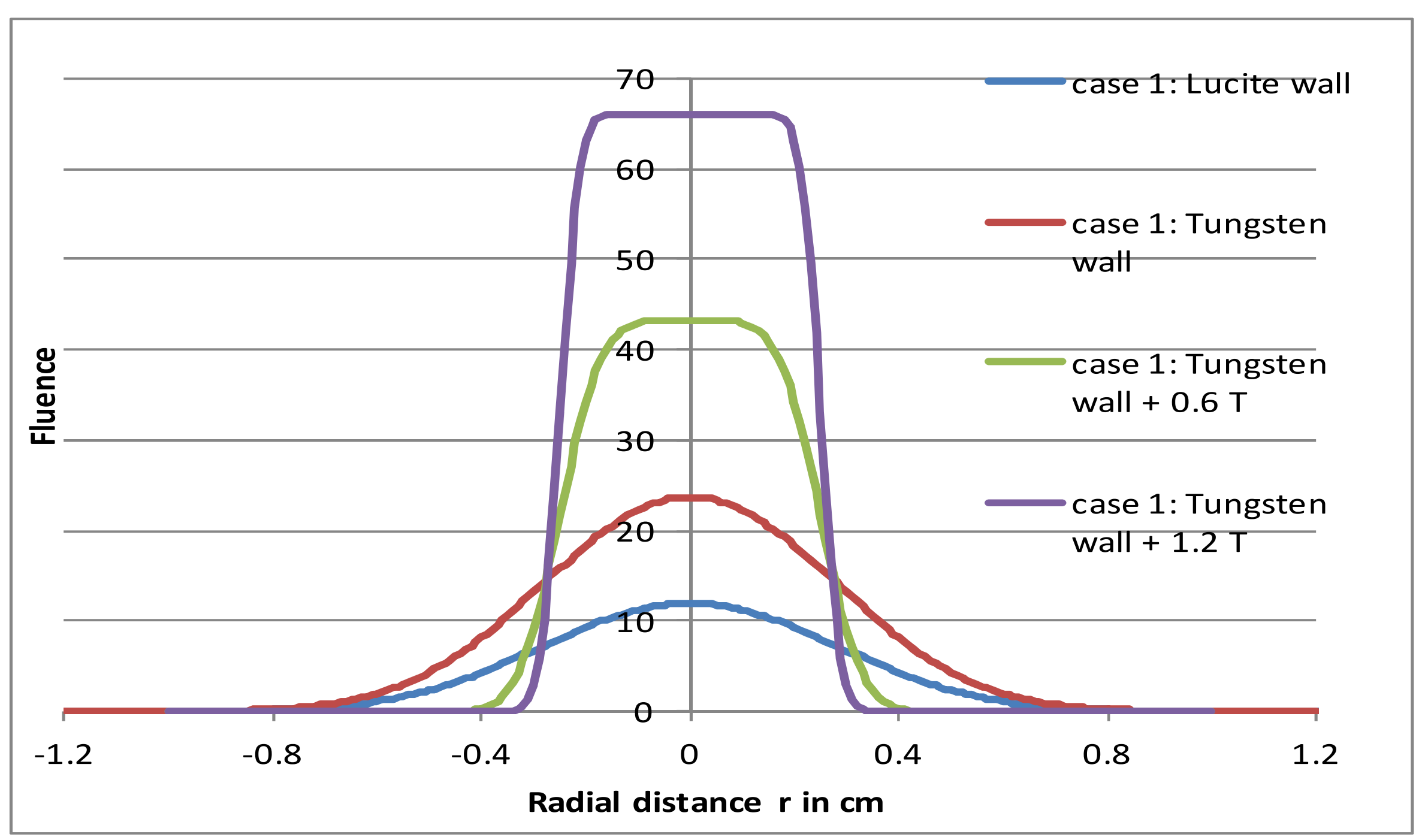

**Figure 14:** Fluence distribution at the end plate of the cone. The cone diameter at this position amounts to 0.5 cm.

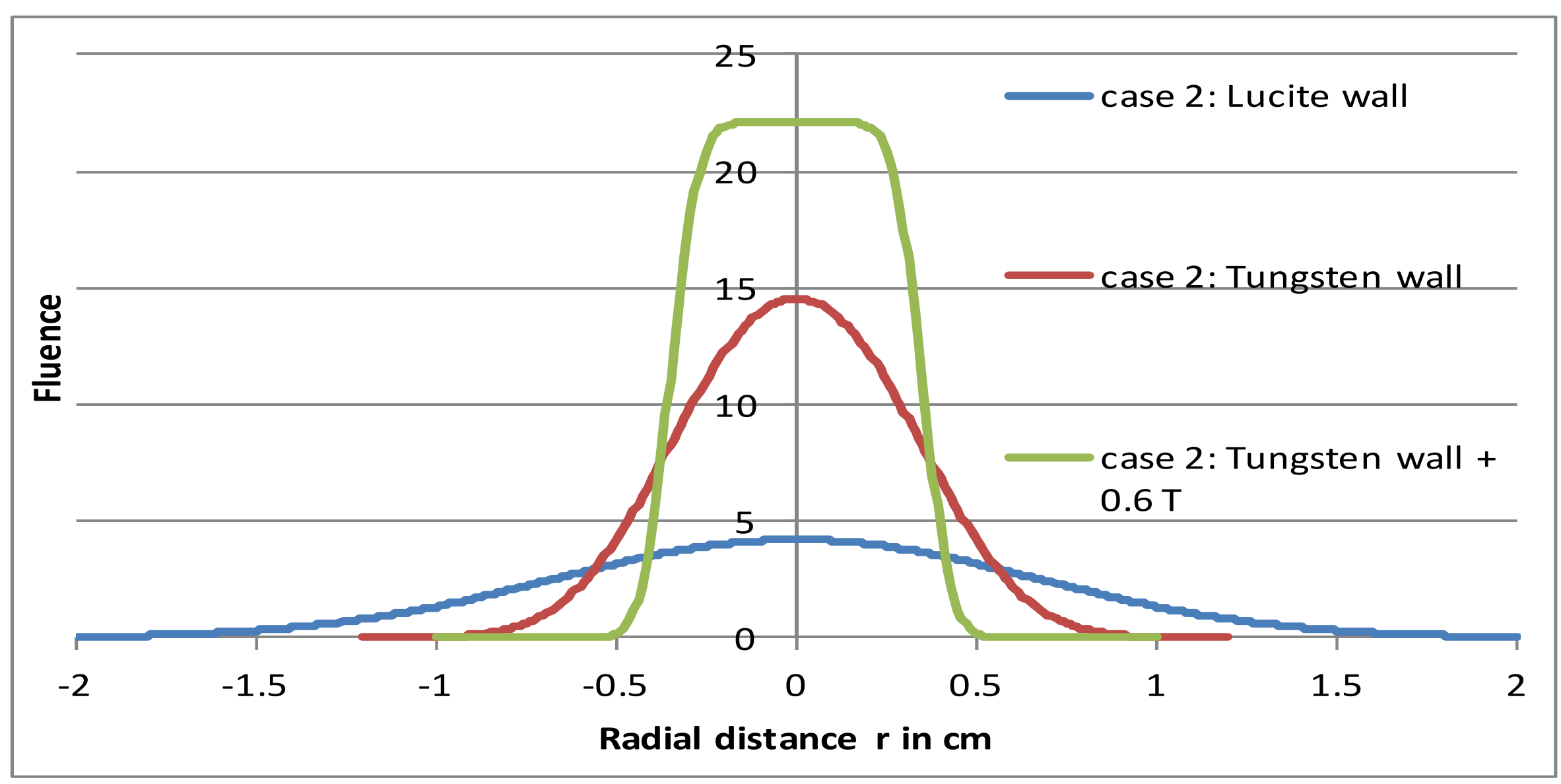

**Figure 15:** The diameter of the cone at the end plate now amounts to 0.7 cm, the case with the magnetic field strength 1.2 T has been omitted, since it is not necessary for increased field sizes.

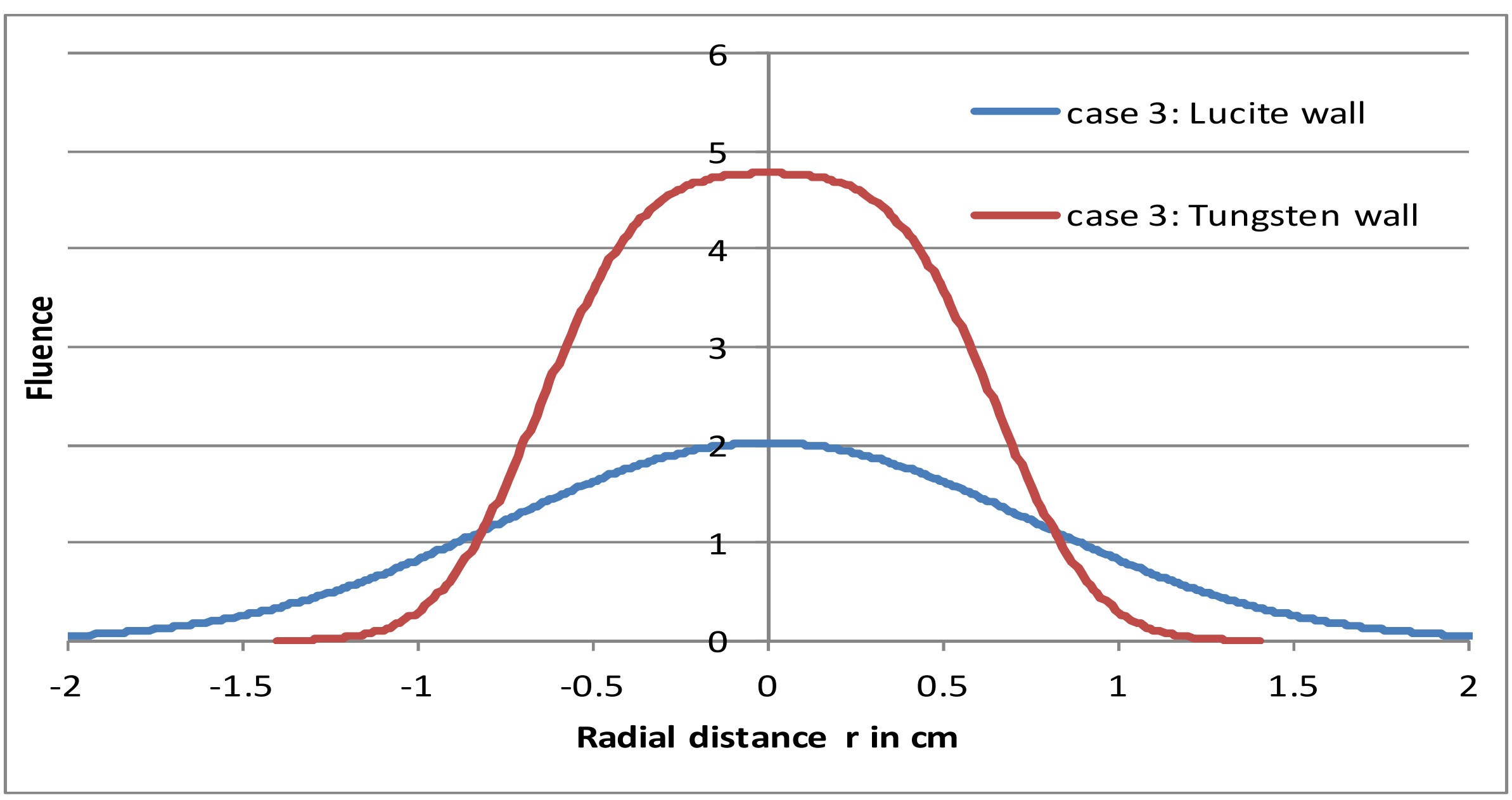

5 **Figure 16:** The diameter of the cone at the end plate now amounts to 1.3 cm (magnetic fields have been omitted).

The following Figures 17 - 22 present the situation at a distance of 90 cm from the end of the cone (calculations and measurements).

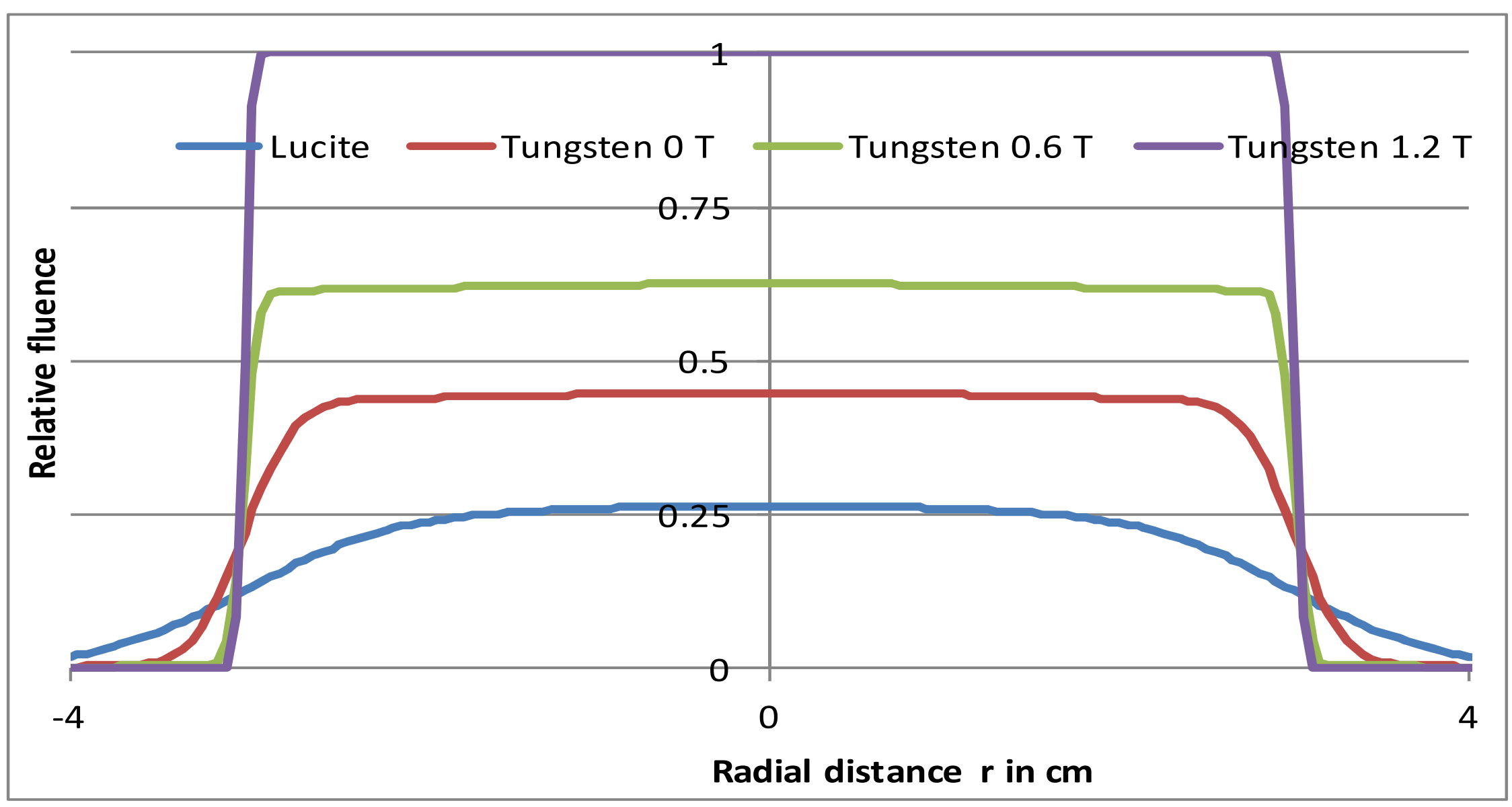

**Figure 17:** Case 1 at z = 90 cm (diameter: 6 cm).

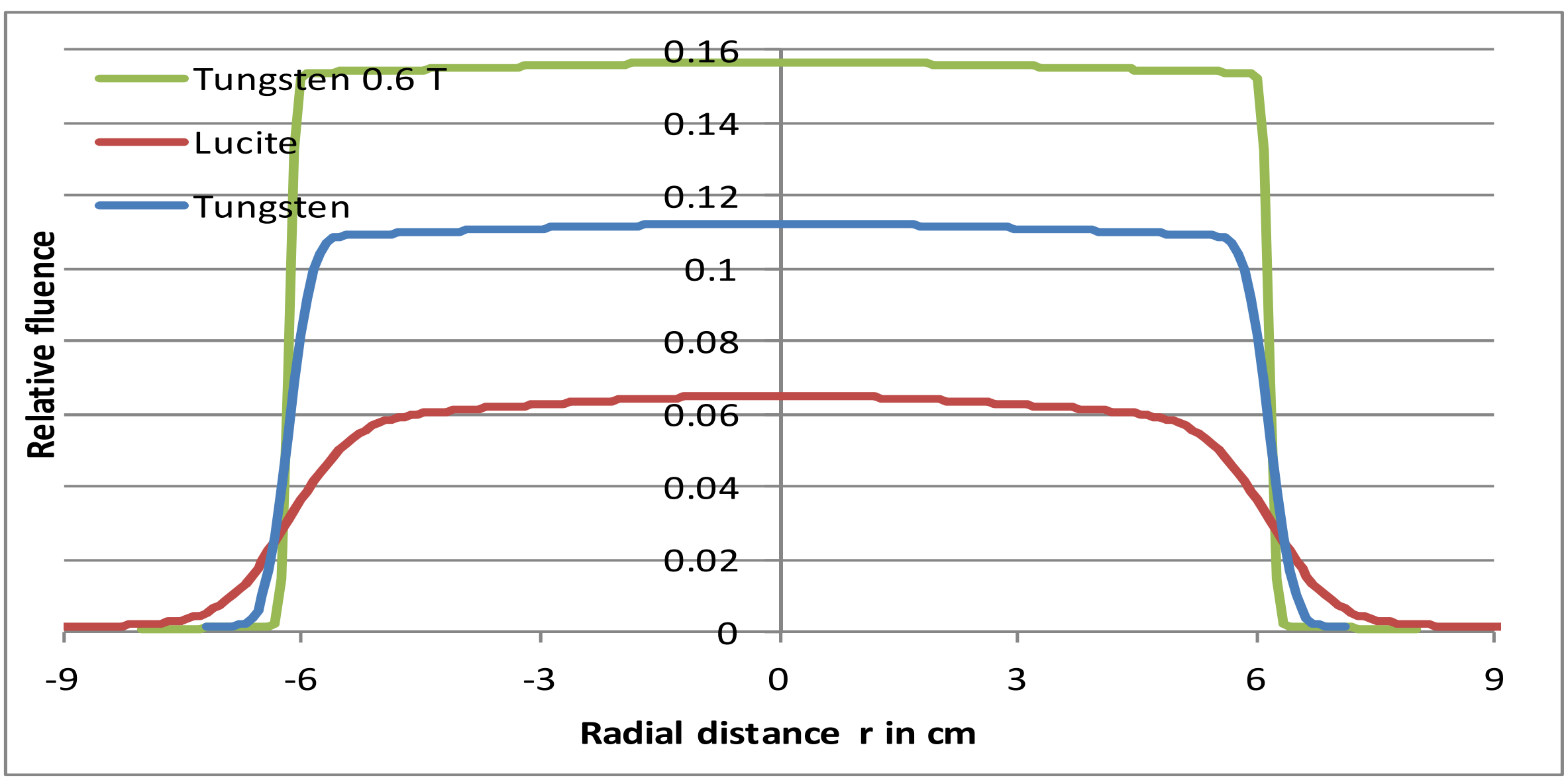

**Figure 18:** Case 2 at z = 90 cm (diameter: 12 cm).

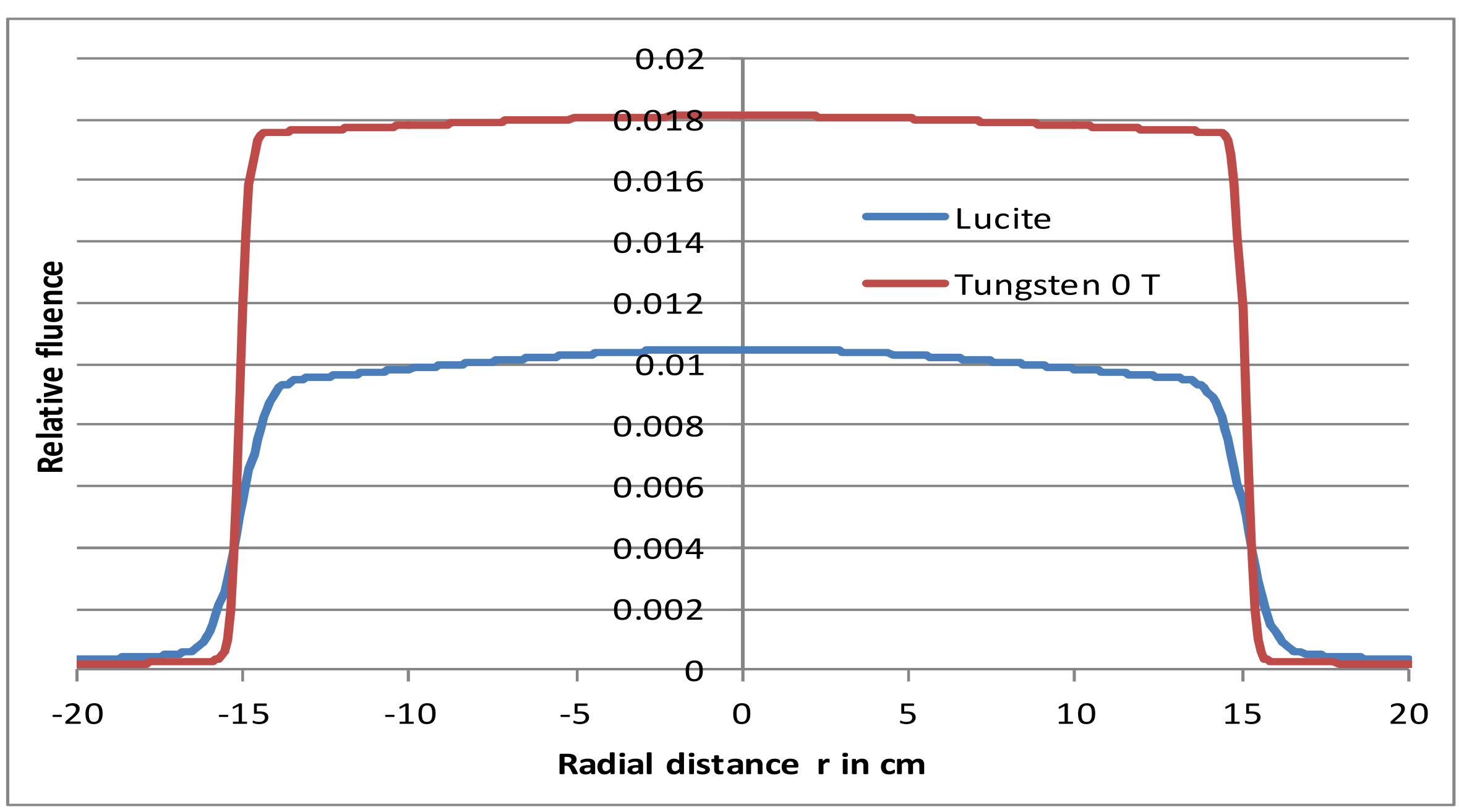

**Figure 19:** Case 3 at z = 90 cm (diameter: 30 cm).

The normalization of the fluence has been taken such that the maximum case according to Figure 17 is '1'.

5 This normalization is also valid with regard to the following Figures 20 – 22 (measurement data).

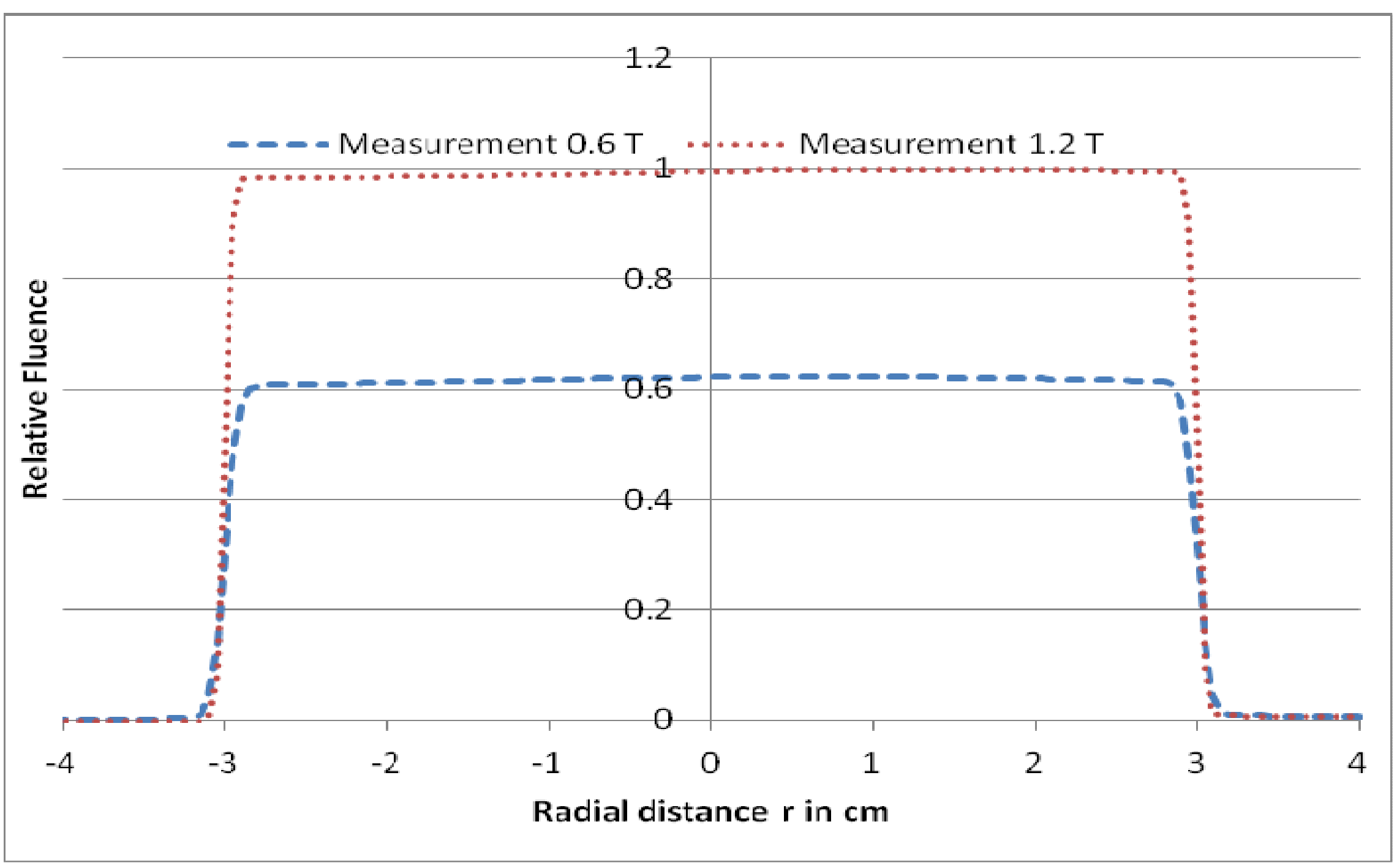

**Figure 20:** Measurement data for the case 1 with 6 cm diameter.

The rather small opening and the focusation by the magnetic fields can obviously compensate the (small) asymmetry of the incoming electron beam. With regard to the larger field sizes the comparison between measurement and calculation appears to be more important.

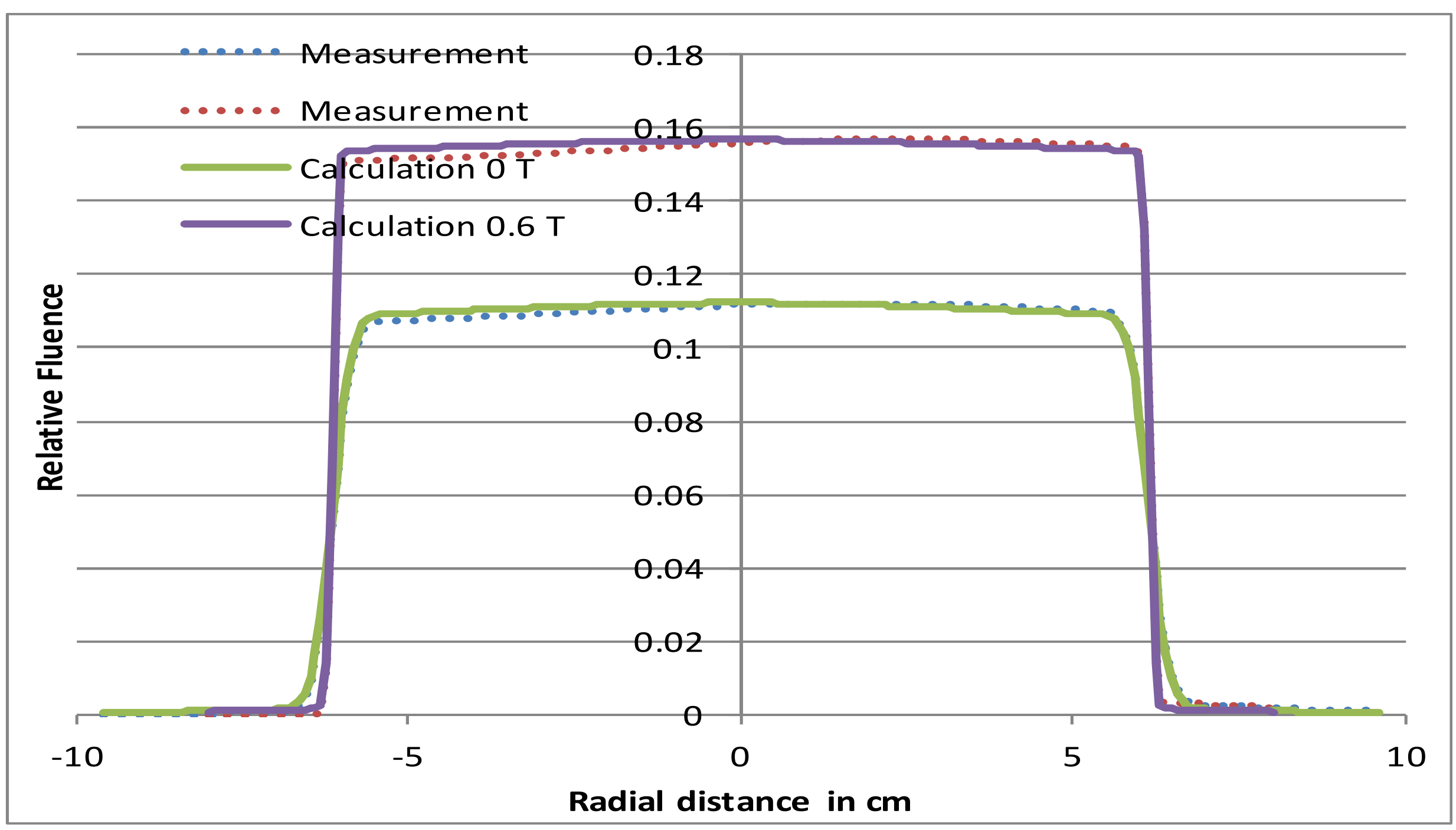

5 **Figure 21:** Calculations and measurements for the case with a field size diameter of 12 cm.

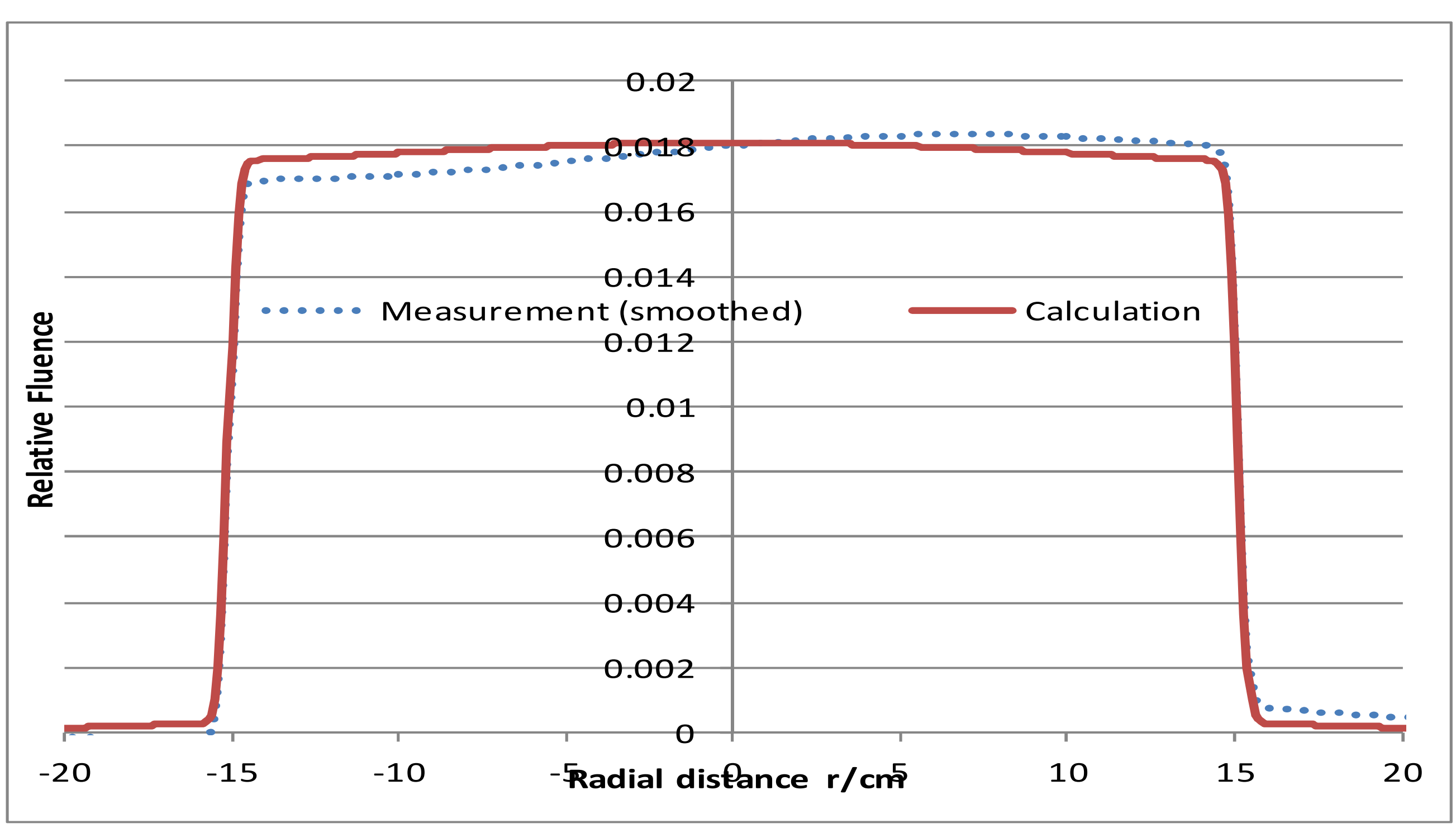

**Figure 22:** Calculations and measurements for the case with a field size diameter of 30 cm.

In particular, the last case shows best the asymmetry in measurements. However, in spit of this fact the agreement between theory noteworthy.

## 4. Discussion and conclusion

5 It could be shown that in conventional linear accelerators used in medicine a multitarget consisting of a Tungsten wall (thickness of the wall at least 2 mm) and 31 very thin plates (thickness of a plate: ca. 0.01 mm) is superior to the standard accelerator. The bremsstrahlung beam (inclusive divergence) can be formed according to the desired properties. The energy spectrum is significantly increasing even in the absence of a focusing magnetic field and is even better than a conventional beam, which has
10 passed a flattening filter. Thus the omission of such a filter provides a further yield of the factor 3 – 4. The optional amplification of the focusing effect by suitable external magnetic fields (with regard to the required properties, see e.g. Figure 6) can be taken into account, in particular, if the outcoming γ-beam should be very efficient by restriction to rather small fields. These properties are important for scanning methods, stereotaxy, IMRT or tomographic applications. It is possible to reach some
15 essential progress in the domain of linear accelerators in radiotherapy, since the modern irradiation techniques such as IMRT, stereotaxy, etc. do not require large field sizes, e.g. a 40 x 40 $cm^2$ at a distance of 100 cm from focus. This progress can be achieved by exploiting small angle reflexion of fast electrons at a Tungsten wall. The wall has to map the desired divergent properties of the beam. *A further aspect of this study is that we are able to save heavy high Z-material for the shielding of the*
20 *accelerator head.* The attached appendix deals with stopping power and heat production of high energy electrons. By that, we have been able to estimate the heat production in each thin plate, which turned out to be lower $20^o$ C per 400 MUs. Thus the systems even works without further cooling of plates, if the rate of MUs will be increased to 1000 or more.

## Appendix: Collision interaction of electrons with matter

The purpose of this appendix is to provide tools for the determination of the heat production of electrons in Tungsten. In a previous publication (Ulmer and Matsinos 2011) we have applied the generalized Bragg-Kleeman rule to proton dosimetry. Therapeutic electrons always satisfy $E_0 >> mc^2 = 0.511$ MeV. An optimum adaptation of relation (A1) to $R_{CSDA}$ of electrons (see Berger et al 2000) shows (Figure A1) that, for $E_0 \leq mc^2$, $p(E_0) \geq 1$; for energies above $mc^2$, $p(E_0) < 1$. For electrons, the factor *A(water) = 0.238552* cm/MeV$^p$ is also rather different to that for protons. The parameters for the calculation of $p(E_0)$ with Formula (1) are given in Table A1.

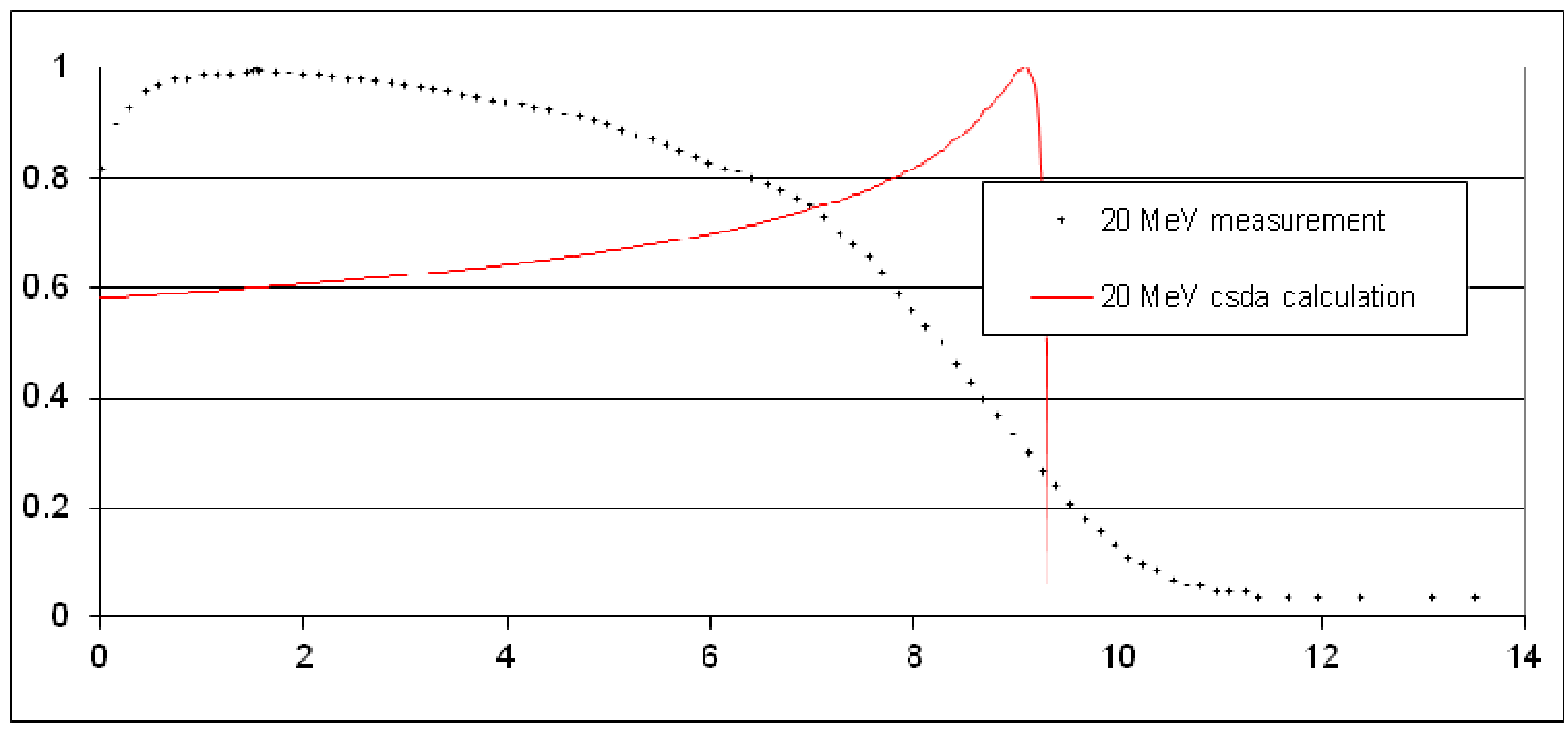

**Figure A1:** 20-MeV electrons and determination of the stopping-power function, obtained in the CSDA framework. The measurements have been obtained in a standard Varian Clinac ('Golden Beam Data').

According to the above citation the Bethe-Bloch equation describing the collision interaction of charged particles can be summarized for electrons by the equation (Ulmer and Matsinos, 2011):

$$\left.\begin{aligned} R_{CSDA} &= A(water)\cdot(E_0 + E_0^2/2mc^2)^p \\ p &= p_0 - c_l E_0/2mc^2 + p_1 \exp(-q_1 E_0/2mc^2) + p_2 \exp(-q_2 E_0/2mc^2) \end{aligned}\right\} (A1)$$

This equation is only valid for water. Since the factor $A$ according to equation (A1) is proportional to $A_w/(\rho_w \cdot Z_w)$, we are able to modify it by the substitution:

$$A(medium) \Rightarrow A(water)\cdot A_m \cdot \rho_w \cdot Z_w/(A_w \cdot \rho_m \cdot Z_m) \quad \text{(A1a)}$$

The meaning of the parameters (water: reference values) of the substitution (A1a) $A_w = 18$, $Z_w = 10$, $\rho_w = 1$ g/cm$^3$, and for other media we have to substitute the corresponding parameters $A_m$, $Z_m$, $\rho_m$ (e.g. Tungsten: $Z_m = 74$, $\rho_m = 19.25$ g/cm$^3$, $A_m = 183.84$)

**Table A1:** The Table values of the dimensionless parameters of Formula (A1)

| $p_0$ | $p_1$ | $p_2$ | $c_l$ | $q_1$ | $q_2$ |
|---|---|---|---|---|---|
| 0.655 | 0.6344 | 0.2616 | 0.0023494 | 3.060 | 0.311 |

The inversion of formula (A1) provides the stopping power S in dependence of the residual energy:

$$S(u)=(R-u)^{1/p-1}/[1+(R-u)^{1/p}/(m_0c^2\cdot A^{1/p})]^{1/2} \quad (A2)$$

$$\left.\begin{aligned} I_B(z)&=I_0\cdot\exp(-\mu z)\\ D(z)&=\int S(u)K(s_0,s_1,u-z)du+I_B(z)\end{aligned}\right\}(A3)$$

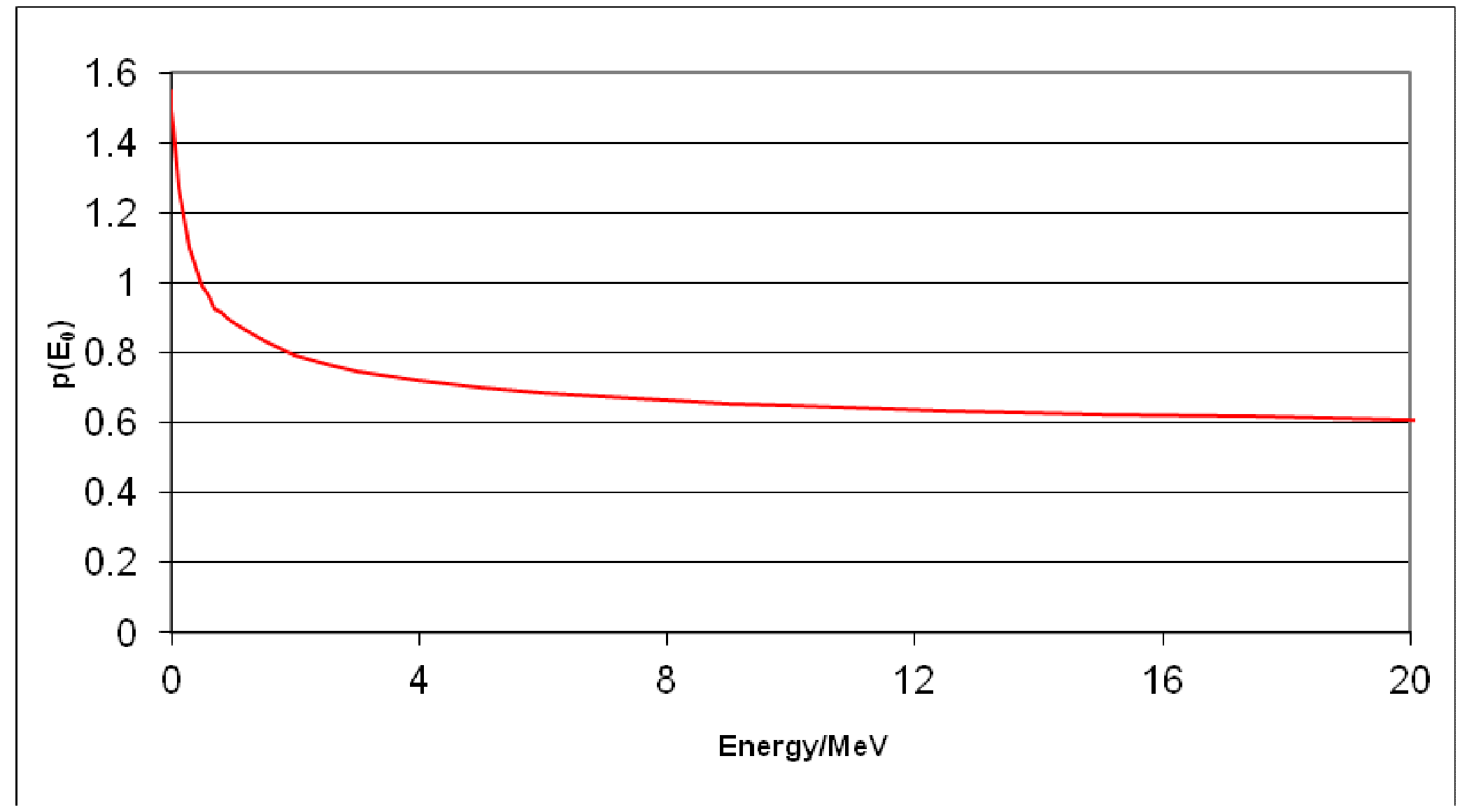

**Figure A2:** Function $p(E_0)$ determined by formula (A1) and ESTAR

The stopping-power formulas (A1 – A3) have to be used for therapeutic electrons; in this case, the length contraction is not a negligible effect. It is also noteworthy that, for $p(E_0) \leq 1$, the singularity of E(s) at s = $R_{CSDA}$ is removed. We have used formula (A3) for the depth-dose calculation of 20-MeV electrons and subjected it to convolutions. The measured and calculated (including the bremsstrahlung effects) curves are shown in Figure A1; the kernel, used in the convolution, and related parameters are displayed in Table A1. The influence of bremsstrahlung is presented in Figure A1 and the modifications, induced by the omission of the effect, can be seen in Figure 44; this figure demonstrates that the parameters of the kernel are rather different from the first case (they are now less plausible), and the tail cannot be described. The CSDA stopping-power is shown in Figure A1. With regard to accounting for bremsstrahlung, we have only considered the contribution resulting basically from the double-scatter foil. This contribution is determined by the software EGSnrc, see Kawrakow and Rogers (2000). The low-energy bremsstrahlung (produced by the electrons in water), its multiple scatter and absorption can be easily explained by the energy-range/straggling (this is basically relativistic and a single Gaussian is not sufficient).